%% file: main.tex
\documentclass[twocolumn,iop]{openjournal}

\input{header/imports}
\input{header/setup}

\input{header/macros}

\begin{document}
\title{JAFF: Bridging astrochemical network formats, live photochemistry and multi-language code generation}

\author{Anish Sarkar
\orcidlink{0009-0006-6879-8751}
}
\email{anish.sarkar1@anu.edu.au}
\affiliation{Research School of Astronomy and Astrophysics, Australian National University, Canberra, ACT 2611, Australia}

\author{James Watt
\orcidlink{0009-0000-0418-8571}
}
\affiliation{Research School of Astronomy and Astrophysics, Australian National University, Canberra, ACT 2611, Australia
}

\author{Mark R. Krumholz
\orcidlink{0000-0003-3893-854X}
}
\affiliation{Research School of Astronomy and Astrophysics, Australian National University, Canberra, ACT 2611, Australia}

\author{
Benjamin Wibking
\orcidlink{0000-0003-3175-2291}
}
\affiliation{The Observatories of the Carnegie Institution for Science, 813 Santa Barbara St., Pasadena, CA 91101, USA}

\author{ChongChong He
\orcidlink{0000-0002-2332-8178}
}
\affiliation{Research School of Astronomy and Astrophysics, Australian National University, Canberra, ACT 2611, Australia}

\author{
Tommaso Grassi
\orcidlink{0000-0002-3019-1077}
}
\affiliation{Center for Astrochemical Studies, Max Planck Institute for Extraterrestrial Physics, 85748 Garching, Germany}

\author{
Piyush Sharda
\orcidlink{0000-0003-3347-7094}
}
\affiliation{Center for Computational Astrophysics, Flatiron Institute, 162 Fifth Avenue, New York, NY 10010, USA}

\author{
Gijs Vermariën
\orcidlink{0000-0002-4346-5858}
}
\affiliation{Leiden Observatory, Leiden University, PO Box 9513, 2300 RA Leiden, The Netherlands}

\author{
Andrea Afruni
\orcidlink{0000-0002-2858-6950}
}
\affiliation{Dipartimento di fisica e astronomia, University of Florence: Florence, Tuscany, IT}

\author{
Alessandro Lupi
\orcidlink{0000-0001-6106-7821}
}
\affiliation{University of Insubria: Varese, Lombardy, IT}

\author{
Brandt Gaches
\orcidlink{0000-0003-4224-6829}
}
\affiliation{University of Duisburg-Essen, Lotharstraße 1, 47057 Duisburg, Germany}

\author{
Munan Gong
\orcidlink{0000-0003-1613-6263}
}
\affiliation{The University of Texas at El Paso, 500 W. University Ave. El Paso, TX 79968}

\author{
Stefano Bovino
\orcidlink{0000-0003-2814-6688}
}
\affiliation{Department of Chemistry, Sapienza University of Rome, P.le Aldo Moro 5, 00185 Rome, Italy}

\begin{abstract}
Chemistry plays an important role in shaping the observable properties of the interstellar and circumgalactic media, as well as protostellar and protoplanetary disks. Chemical processes in these environments are often out of equilibrium, so numerical simulations with detailed chemistry are necessary both for accurate physical predictions and for the validation of simulations against observations. However, current numerical infrastructure for incorporating on-the-fly chemistry into numerical simulations is limited by two factors: lack of cross-compatibility of chemical reaction representation formats across multiple codebases, which forces slow and error-prone re-implementation of chemical networks across every hydrodynamics code that wishes to make use of them, and a lack of support for self-consistent photochemistry, where chemical processes are coupled with radiative transfer as well as hydrodynamics. To address these shortcomings we present \jaff, a cross-platform framework capable of parsing multiple chemical network formats, generating implementations of these networks in multiple languages and in a format suitable for including into any large hydrodynamics code, and with support for self-consistent treatments of photochemistry that couple chemistry and multi-frequency radiative transfer, in addition to the simple effective-extinction approximations frequently adopted in earlier work. We present a test suite demonstrating the use of \jaff-generated code to add both pure chemistry and photochemistry networks to multiple large hydrodynamics codes. The \jaff framework is available at \href{https://github.com/jaff-chemistry/jaff}{https://github.com/jaff-chemistry/jaff} and detailed documentation is available at \href{https://jaff-chemistry.github.io/jaff/}{https://jaff-chemistry.github.io/jaff/}.
\end{abstract}


\maketitle

\section{Introduction}
\label{sec:intro}

Much of our knowledge of the cool and cold cosmos comes from observing the radiation emitted or absorbed by molecules. By acting as sources of cooling, these molecules also play an important role in the evolution of astrophysical sources such as the interstellar medium \citep[ISM,][]{vdb88, van_dishoeck_blake-98,tielens-05,draine-ism-11}, protostellar cores \citep{bergin-tafalla-2007,caselli-caccarelli-12,aikawa+2026}, protoplanetary disks \citep{armitage-2011,henning-semenov-13,andrews-2020,oberg-bergin-21}, and galactic winds \citep{Veilleux20a,Di-Teodoro20a, Vijayan24b, noon-26}. In systems where chemical processes are important to physical evolution, or where the chemical state of the gas is far from equilibrium, simulations must treat the dynamics and chemistry as a coupled numerical problem where both processes are solved simultaneously. The degree of complexity with which these processes are treated varies considerably between numerical models, reflecting differences in their scientific objectives, physical assumptions, and computational requirements. For example, simulations may adopt different levels of chemical detail, ranging from reduced networks designed to capture the dominant chemical pathways to extensive networks incorporating hundreds of species and thousands of reactions. Similarly, the treatment of the gas dynamics and associated microphysics can range from simplified prescriptions to fully three-dimensional, self-consistent calculations. In either approach, numerical chemical reaction networks (CRNs) are the key tool to do chemistry. A CRN is a topological representation of the chemical processes occurring in a medium, which combines a list of chemical species tracked with a set of reactions describing the rates at which the abundances of these species change; the sum of these reactions then defines the right-hand side (RHS) of a system of non-linear ordinary differential equations (ODEs) that must be solved together with the (radiation- and magneto-)hydrodynamic evolution of the system. 

Over the years, a number of databases and corresponding data formats have been introduced to represent CRNs, and some of these have been coupled with software to automatically generate code representing the RHS in a format suitable for inclusion in a larger hydrodynamic simulation code \citep{kida,uclchem,krome,prizmo,udfa}. However, existing software for this purpose lacks cross-platform and cross-language support, and as a result is often limited to one or a handful of simulation codes. For example, the widely-used KROME package \citep{krome} only supports CPU based codes, among which \code{flash} \citep{flash}, \code{enzo} \citep{enzo}, \code{gadget} \citep{gadget} and its siblings, \code{ramses} \citep{ramses}; the PRIZMO package supports \code{pluto} \citep{pluto}; \code{arepo} and \code{athena++} have their own dedicated chemistry code-generation packages \citep{hg15, Gong23a}. While there are packages for network analysis and modeling \citep[e.g.~\code{catalyst},][]{Catalyst-jl}, at present, no general-purpose package exists for unifying CRN software or formats. Although the new \code{krome-x} \citep{krome-x} package  facilitates interfacing with a broader range of codebases, including GPU based codes, through its \texttt{skill.md} file, it still relies on its own custom CRN representation format. This has several important negative consequences.

First, this makes code comparison extremely difficult, since if every code uses its own single-purpose CRN-generator then there is no simple way to guarantee that two hydrodynamics codes are actually solving the same chemistry network. Second, it leaves many widely-used hydrodynamics codes -- e.g.,  \code{athena++} \citep{athenapp} -- with no publicly available automated facilities to translate CRNs into executable code. Third, at present most of the publicly-available code-generation packages do not support GPU acceleration, as would be required for the resulting code to be usable in GPU-oriented hydrodynamics and chemistry codes such as \code{quokka} \citep{quokka}, \code{gpluto} \citep{gpluto}, \code{athenaK}/\code{k-athena} \citep{Stone2026_athenaK,kathena}, \code{gamer}, and \code{gamer} \citep{gamer}, the only exception being \code{krome-x} \citep{krome-x}. Of course, one can work around all of these problems by hand-writing an implementation of a particular CRN for a particular code, but since many CRNs contain thousands of reactions \citep[e.g.,][]{kida,kida_uva_2024,udfa}, this is impractical and error-prone.

A further limitation in existing CRN software has also become more pressing as
growth in computational power and resources has allowed the simulation of increasingly complicated physical processes with additional self-consistent physics. 
Much of interstellar chemistry is radiation-driven, but due to computational cost in the past the most common approach to this has been to use shielding approximations to estimate the radiation field at a given position \citep[e.g.,][]{Clark12a, Hartwig15b, Safranek-Shrader17a}. In recent years, there has been a shift to tackling self-consistent\footnote{In our discussions and examples, we use the phrase ``self-consistent'' to mean that the time-dependent radiation field evolved along with the chemistry, rather than a using a fixed radiation field or a prescription that derives the radiation field from a prescription based on gas properties. This does not imply a self-consistent solution obtained from the level populations of individual species.} approaches where one solves the chemical evolution equations simultaneously with the radiative transfer equations in much the same way that earlier approaches solved chemistry simultaneously with the equations of hydrodynamics or MHD \citep[e.g.,][]{glover2010,lupi+18,lupi+2020,lupi+2021,Katz22b, Gong23a, Kim2023_photochemistry,Peter23a,lupi+2024, Peter25a,popsicle,2026arXiv261017577M, Ho2026}, but at present this effort relies on CRNs (often manually implemented) that are locked to one particular code and on a particular set of chemical reactions. As with standalone CRNs that do not include radiation, this approach is error-prone, inhibits code comparison, and limits extensibility. All these issues warrant the necessity of an automated, battle tested, extensible pipeline capable of parsing networks and generating code for both pure chemistry and photochemistry.

In this work, we present \jaff (Just Another Fancy Format), a new extensible and future-proof cross-platform framework for CRNs. At its core, \jaff is a plugin-based parser which reads all major CRN formats -- \code{prizmo} \citep{prizmo}, \code{krome} \citep{krome}, \code{kida} \citep{kida,kida_uva_2024}, \code{udfa} \citep{udfa} and \code{uclchem}\footnote{Support for the new \code{krome-x} format will be added in a future release.} \citep{uclchem,uclchem-4.0} -- and validates each reaction. Parsed networks are stored as a symbolic graph (powered by \code{sympy}) which can generate network, reaction and species statistics with support for further analysis and visualization. \jaff further supports a templated, API-free code generator capable of emitting expressions in seven languages -- C, C++, Fortran, Python, Julia, Rust and R -- from a single command; complete templates and solvers are provided for C, C++, Fortran, Python and Julia, with GPU decorators where GPU support is required. Finally and most importantly, \jaff offers a self-consistent treatment of photochemistry and multi-band radiation, which is not locked to a particular method of handling radiation fields, and can instead provide drop-in support for photochemistry in any code that includes radiative transfer.

The remainder of this paper is organized as follows. \Cref{sec:framework} describes the \jaff framework and its parsing, thermodynamics, code generation and photochemistry capabilities. \Cref{sec:test-suite} validates the generated code against analytical, equilibrium and radiation-coupled benchmarks across multiple simulation codes. We summarize in \Cref{sec:summary} and discuss limitations and future work.

\section{The \jaff framework}
\label{sec:framework}

Written in Python and designed for flexibility and extensibility, \jaff exploits the symbolic capabilities of \code{sympy} to parse and evaluate reaction rates. 
Because the rates are stored symbolically, the same model doubles as an object-oriented CRN analysis tool with capabilities such as reaction, species and element statistics, rate and cross-section visualization, table inter-conversions, and code generation. Runtime parameters are supplied through a \textit{.toml} configuration file, so a network's behavior is preserved between runs. Two command-line tools ship alongside the library -- \jaffx for quick reaction-metadata extraction and \jaffgen for code generation. Here, we summarize the major features of \jaff, with particular attention to its self-consistent treatment of multi-band photochemistry and dust extinction.

\subsection{The core architecture}
\label{ssec:internal-architecture}
\jaff's core is an abstract representation of CRNs, implemented as a Python class \code{network}, the structure of which we illustrate in \Cref{fig:network-arch}. This class stores all the properties of a CRN in the form of objects and attributes, including a list of all the elements and chemical species included in the network, a list of reactions involving those species, and the total rate of internal energy change due to both chemical reactions and non-chemical sources (e.g., collisionally-excited line emission not associated with a chemical reaction). Reactions are Python classes that store an abstract representation of a reaction occurring in the CRN. The stored information includes all elements, species, reactants and products of the reaction, with corresponding stoichiometric factors, any restriction of the temperature range over which the reaction occurs, the change in internal energy per reaction, the reaction rate, and the reaction type (e.g., photo-reaction, two-body reaction, three-body reaction, dust-assisted reaction). The reactions together determine the evolution equations for the species abundances and the gas internal energy for that network; specifically, the rate of change for the number densities and gas volumetric internal energy density $E$ obey
\begin{eqnarray}
    \dv{n_i}{t} & = & \sum_{j} (\beta_{ij}^{\rm products} - \beta_{ij}^{\rm reactants}) \mathcal{R}_j
    \label{eq:general-rates}\\
    \dv{E}{t} & = & \sum_j \Delta E_j \mathcal{R}_j + \dot{E}_{\rm nc}
    \label{eq:general-edot}
\end{eqnarray}
where $n_i$ is the number density of species i, $\beta_{ij}^{\rm products}$ and $\beta_{ij}^{\rm reactants}$ are the dimensionless stoichiometric coefficients on the product and reactant sides, respectively, in reaction $j$, $\mathcal{R}_j$ is the reaction rate per unit volume for reaction $j$, $\Delta E_j$ is the change in gas thermal energy per reaction $j$, and $\dot{E}_{\rm nc}$ is the rate of change of the gas internal energy per unit volume due to non-chemical processes. For a typical two-body reaction of the form $aR_1 + bR_2 \rightarrow \ldots$, where $R_1, R_2$ denote the reactants and $a, b$ their corresponding stoichiometric coefficients, $\mathcal{R}$ is given by the law of mass action 
\begin{equation}
    \label{eq:2-body-reaction}
    \mathcal{R} = n_{R_1}^{a}n_{R_2}^{b}k
\end{equation}
where $n_X$ denotes the number density of species $X$ and $k$ is the reaction rate coefficient. However, reactions are not required to follow this format -- indeed, a crucial feature of \jaff is that the reaction rates $\mathcal{R}_j$, energy changes $\Delta E_j$, and non-chemical energy density change rate $\dot{E}_{\rm nc}$ are not restricted to being single numbers or formulae. Instead, they are \code{sympy} expressions that can encode arbitrary functions of the state of the system, and can include tabulated, interpolated data (e.g., tables of reaction rate as a function of temperature) as well as analytic expressions. Consequently, while \jaff provides the ability to generate these \code{sympy} representations automatically from standard CRN formats that assume reaction rates to follow a standard functional dependence on reactant density and temperature (as we discuss below in \Cref{ssec:multi-format}) or from frequency-dependent cross-sections and radiation fields (as we discuss in the next section), it is not restricted to these particular functional forms. Instead, reaction rates and chemical and non-chemical energy changes can depend on arbitrary properties of the system, and \jaff provides a standard format for specifying such arbitrary dependencies (introduced in \Cref{ssec:custom-rates}). For example, they can depend on the ambient radiation field -- as required for reactions where the mean grain charge state (which is affected by the radiation field) matters, or they can depend on total gas density and ionization state -- as required to handle cosmic ray-driven reactions where the amount of thermal energy deposited in the gas depends on its density and chemical state \citep{Glassgold12a}. 

For networks that include photochemistry, reactions also carry information about how the reaction interacts with the radiation field, as we detail in the next section.
\input{figure_templates/arch}
\subsection{Photochemistry}
\label{ssec:photochem}

The primary goal of \jaff's handling of photochemistry is to provide self-consistent evolution equations for both the chemical abundances and the radiation field coupled to a numerical implementation. As with all other reaction rates, \jaff allows rates, energy changes, and changes in the radiation field for photochemical reactions to be arbitrary \code{sympy} expressions. However, it also provides the ability to generate reaction rate expressions automatically for arbitrary user-specified discretizations of the radiation field in photon energy; we describe this automatic generation pathway in the remainder of this section.

\subsubsection{Reaction rate and photo-heating expressions}
\label{sssec:photo-reaction-rates}

The automatic rate generation pathway in \jaff~assumes that the reaction rate for a given photon-driven reaction
\begin{equation}
    \label{eq:photo_reaction}
    A+\text{h}\nu\rightarrow  {\rm products}
\end{equation}
can be expressed in the generic form
\begin{equation}
    \label{eq:photo_reaction_rate}
    \mathcal{R} = n_Ak
\end{equation}
where $n_A$ is the number density of species $A$, and where the rate coefficient $k$ is given by
\begin{equation}
    \label{eq:rad_reaction_rate_coefficient}
    k = \hat{c}\int_{E_{\rm th}}^{\infty}\sigma_r(E_{\gamma}) \dv{n_\gamma}{E_\gamma}\dd{E_\gamma}.
\end{equation}
Here $E_{\rm th}$ is the activation energy of the photo reaction, $\hat{c}$ is the reduced speed of light\footnote{\jaff supports the reduced speed of light approximation \citep[e.g.,][]{Gnedin16a}, whereby reaction rates are multiplied by a reduced speed of light value $\hat{c}$ in place of the full speed of light $c$. However, by default \jaff uses $\hat{c} = c$.}, $\sigma_r(E_\gamma)$ is the cross-section for the reaction (a topic we defer to \Cref{sssec:photo-reaction-cross-section}), and $\dd{n_\gamma}/\dd{E_\gamma}$ is the number density of photons per unit photon energy $E_\gamma$.

\input{figure_templates/oh_cross_section}

Like \code{krome}, \jaff's reaction generation machinery assumes that the radiation field is discretized into distinct bands $b$, defined by minimum and maximum energies $(E_{\gamma,b-/2}, E_{\gamma,b+1/2})$; the discretized quantity can be either the photon number density in each band, $n_{\gamma,b} = \int_b (\dd{n_\gamma}/\dd{E_\gamma}) \, \dd{E_\gamma}$, or the radiation energy density in the band, $u_{\gamma,b} = \int_b E_\gamma (\dd{n_\gamma}/\dd{E_\gamma}) \, \dd{E_\gamma}$, where we use the shorthand that $\int_b$ means integration over the photon energies in the band, i.e., over the range $(E_{\gamma,b-1/2}, E_{\gamma,b+1/2})$. Note that the choice of discretization -- both in terms of the placement of the band edges and whether the discretization is done in terms of photon number or energy -- need not be specified when a network is created in \jaff or when a CRN is written to disk, as all of the calculations below are done on-the-fly as needed, rather than being pre-computed.

For a discretization by number, we can define a band-averaged cross-section by
\begin{equation}
    \langle\sigma\rangle_{r,b} = \frac{\int_b \sigma_r(E_\gamma)\mathcal{W}(E_\gamma) \dd{E_\gamma}}{\int_b \mathcal{W}(E_\gamma)\dd{E_\gamma}},
    \label{eq:mean-photo-xsec}
\end{equation}
where $\mathcal{W}(E_\gamma)$ is the assumed spectral shape $\dd{n_\gamma}/\dd{E_\gamma}$ within the frequency band; \jaff defaults to assuming $\mathcal{W}(E_\gamma) \propto E_\gamma^{-2}$, corresponding to constant photon energy per logarithmic bin in photon energy\footnote{We note that an alternative approach would be to construct the shape from neighboring bands in real-time. However, a number of limitations are imposed in the optically thick limit, wherein the local radiation energy spectrum must approach the Planck spectrum and the local radiation flux must tend to zero. Under such conditions, assuming a spectral slope of $\mathcal{W}(E_\gamma) \propto E_\gamma^{-2}$ proves to be the most robust choice. We refer the reader to Sections 2.2.2.1 and 2.2.2.2 of \citet{quokka-multigroup} for details.} for all bands, but users can override this choice by supplying a separate power-law index and thus a different spectral shape for each band. Given this definition, we can express the reaction rate due to photons in band $b$ as
\begin{equation}
    \mathcal{R}_b = \hat{c} n_A n_{\gamma,b}\langle \sigma\rangle_{r,b}
    \label{eq:photo-reaction-rate-approximation}
\end{equation}
where $n_{\gamma,b}$ is the photon number density in band $b$; the total reaction rate is $\mathcal{R} = \sum_b \mathcal{R}_b$, and \jaff~adds this term to the right-hand side of the species evolution equations, \Cref{eq:general-rates}, for each photoreaction. The rate per unit volume at which the photons deposit energy in the gas is then
\begin{equation}
    \dot{u} = \sum_b \mathcal{R}_b \left\langle E_{\gamma}\right\rangle_b,
\end{equation}
where
\begin{equation}
    \langle E_\gamma\rangle_{b} = 
    \frac{\int_b E_\gamma \mathcal{W}(E_\gamma) \dd{E_\gamma}}{\int_b \mathcal{W}(E_\gamma)\dd{E_\gamma}}
\end{equation}
is the band-averaged photon energy for the assumed spectral shape. However, at present \jaff~does not automatically add a corresponding term in the energy evolution equation, \Cref{eq:general-edot}, because for many photoreactions only a portion of the photon energy (less the change in chemical energy) is ultimately deposited as gas thermal energy, while the rest is re-radiated \citep[e.g.,][]{Glassgold12a}. The partition between internal energy and re-radiation can be supplied through user-defined functions; see \Cref{ssec:custom-rates} for details.\footnote{In a future release we intend to add functions to supply the correct expressions for the energy deposited for common photoreactions such ionization of atomic and molecular hydrogen.}

For the case of an energy discretization, \jaff~first computes the rate of energy absorption from each band as
\begin{equation}
    \dot{u}_b = \hat{c} n_A  u_{\gamma,b} \langle \sigma  \rangle_{E,r,b}
    \label{eq:band_energy_deposition}
\end{equation}
where
\begin{equation}
    \langle \sigma \rangle_{E,r,b} = \frac{\int_b \sigma_r(E_\gamma) E_\gamma \mathcal{W}(E_\gamma) \, \dd E_\gamma}{\int_b E_\gamma \mathcal{W}(E_\gamma) \, \dd E_\gamma}
    \label{eq:mean-photo-xsec-energy}
\end{equation}
is the energy-averaged cross section (c.f.~\autoref{eq:mean-photo-xsec}, which is the number-averaged cross section). The reaction rate from the band is then given by
\begin{equation}
    \mathcal{R}_b = \frac{\dot{u}_{b}}{\left\langle E_\gamma\right\rangle}_b.
\end{equation}
The difference between the number-based and energy-based discretization is simply a matter of which quantity one wishes to conserve exactly: the number-based one conserved photon number exactly but radiation energy only approximately, and the energy-based one does the opposite.

We note that both the number-weighted and energy-weighted mean cross sections are only approximations to the actual cross-section at a given frequency for broad bands, and that if band edges are not chosen to coincide with activation energies then photons with zero cross-sections can still cause photo-reactions; it is incumbent on users to choose sensible frequency discretizations for their chosen chemistry network, and \jaff~simply uses the frequency discretization that the user supplies.

We show an example of \jaff's automated computation of mean cross sections, for the case of number-weighed means, in \Cref{fig:oh_xsec}; in this example, the energy range from 5.5 to 14.5 eV has been discretized into 2, 4, and 8 energy bands, and we have used \jaff~to generate band-averaged cross-sections for the reaction $\mathrm{OH} + \gamma\to\mathrm{O} + \mathrm{H}$, using a frequency-dependent cross-section taken from the Leiden database \citep[see \autoref{sssec:photo-reaction-cross-section} for details]{hrodmarsson-2023}. The case in which the radiation frequency has been discretized into eight bands is, not surprisingly, the closet estimate of the actual cross-section relative to the two and four band discretizations. The effect of this approximation is further discussed in our tests in \Cref{sssec:gow-1d-pdr-lpc}.

%
%
%
%

%
%

\subsubsection{Evolution equations for the radiation field}
\label{sssec:moment-equations}

As noted above, one of the main goals of \jaff's treatment of photochemistry is to automate generation of evolution equations not just for chemical species, but also for the radiation field, and to do so in such a way as to guarantee photon or energy conservation to machine precision for situations where conservation is physically expected. Thus whenever photochemistry is enabled, \jaff also generates expressions to describe the evolution of the radiation bands. 

For a photon number-based discretization, \jaff's default expression for the rate of change of the photon density in a radiation band is
\begin{equation}
    \pdv{n_{\gamma,b}}{t} = -\sum_i \mathcal{R}_{i,b},
    \label{eq:rad-equation-1}
\end{equation}
where the sum is over all photoreactions $i$ and $\mathcal{R}_{i,b}$ is the expression given by \autoref{eq:photo-reaction-rate-approximation} for each reaction and band. For a discretization that follows radiation energy density rather than photon number density, the corresponding expression is
\begin{equation}
    \pdv{u_{\gamma,b}}{t} = -\sum_i \dot{u}_{i,b},
    \label{eq:rad-equation-u-1}
\end{equation}
where $\dot{u}_{i,b}$ is the expression given by \autoref{eq:band_energy_deposition} for each reaction and band. Either of these expressions can also be modified in several ways.

First, for some photoreactions there is a distinction between the reaction cross-section $\sigma_{r}(E_\gamma)$ and the absorption cross-section $\sigma_a(E_\gamma)$; a prominent example of this is H$_2$ photodissociation, where only $\approx 15\%$ of absorptions result in a dissociation \citep[e.g.,][]{db96}. In situations where $\sigma_r \neq \sigma_a$, \jaff computes the chemical reaction rates $\mathcal{R}_{i,b}$ in the expression above using $\sigma_a$, thereby accounting for photons that are absorbed but do not induce reactions. Note that \jaff~does \textit{not} automatically add corresponding source terms to account for the fate of photons that are absorbed but do not cause reactions, because this is highly dependent on nature of the reaction being modeled, and possibly also the physical conditions in the gas -- in the case of H$_2$, the absorbed energy is mostly reprocessed into lower-energy photons, but the frequency distribution of the re-radiated emission depends on the density and dust abundance in the absorbing gas \citep{Sternberg88a}. The spectral yield functions described in \Cref{ssec:custom-rates} provide a mechanism by which users can specify these rates and thereby enforce energy conservation for absorptions, but \jaff~does not automate this process because doing so would require adopting some particular set of assumptions about the ultimate fate of the energy carried by the absorbed photons.

Second, reactions in CRNs can be declared as photon-producing, by supplying a rate $\dot{n}_{i,b}$ at which this reaction produces photons in each band for a number discretization, or equivalently a rate $\dot{u}_{i,b}$ for an energy discretization. Thus, for example the hydrogen recombination reaction (${\rm H}^+ + e^- \to {\rm H}$) can be declared to produce photons in a non-ionizing band at a rate $\alpha_B n_{e^-} n_{\rm H^+}$ and in an ionizing band at a rate $(\alpha_A - \alpha_B) n_{e^-} n_{\rm H^+}$, where $\alpha_A$ and $\alpha_B$ are the case A and case B recombination rate coefficients. 

Finally, for two-moment radiation methods (e.g., variable Eddington tensor or M1 methods), \Cref{eq:rad-equation-1} or \Cref{eq:rad-equation-u-1} can be supplemented by a second evolution equation for the radiation flux, either
\begin{equation}
    \pdv{f_{\gamma,b}}{t} = -n_A\hat{c}\sum_i f_{\gamma,b}\langle\sigma\rangle_{r,b},
    \label{eq:rad-equation-2}
\end{equation}
for a photon number flux $f_{\gamma,b}$ or
\begin{equation}
    \pdv{F_{\gamma,b}}{t} = -n_A\hat{c}\sum_i F_{\gamma,b}\langle\sigma\rangle_{E,r,b},
    \label{eq:rad-equation-u-2}
\end{equation}
for a radiation energy flux $F_{\gamma,b}$. Note that photon-producing reactions are assumed \textit{not} to contribute to flux evolution, which is equivalent to assuming that photons are emitted isotropically and thus carry zero net flux.

\subsubsection{Shielding functions}
\label{sssec:shielding}

The treatment of reaction rates and radiation fields described thus far is sufficient to handle reactions where the energy-dependent cross-section $\sigma_r(E_\gamma)$ can reasonably be approximated as a band average, which is the case when the cross-section is smooth except for the presence of a small number of edges; thus this approximation is reasonable for almost all ionization reactions, for example ionization of atomic hydrogen and carbon. However, replacing $\sigma_r(E_\gamma)$ by a small number of band-averages $\langle\sigma\rangle_{r,b}$ is a poor approximation for reactions driven by resonant absorptions for which the cross-section has complex energy dependence, for example H$_2$ photodissociation by resonant absorption of Lyman-Werner band photons. In such cases the band-averaged approximation is poor unless the bands are finely-sampled enough to resolve the structure of $\sigma_r(E_\gamma)$, which in practice is almost never the case for three-dimensional simulations except for very specific cases \citep[e.g.][]{kristensen+2023}. Instead, the usual approach to handling reactions of this type is through empirical shielding functions that approximately capture the key physical process of a reduction in the effective reaction and absorptions rates produced by radiation fields that have already traversed a significant column of absorbing molecules due to selective depletion of photons at energies near resonance \citep[e.g.][]{db96,vvdb09,wolcott-green+2011}. 

\jaff supports this shielding function approach as follows: \jaff allows each photoreaction to have an associated shielding function $f_{s,i}$ that approximates the effects of shielding -- i.e., reduction in the reaction rate due to selective depletion of resonant photons -- by species $i$. For reactions to which a shielding function applies,  \Cref{eq:rad-equation-1,eq:rad-equation-2} are modified by replacing the full cross section $\langle\sigma\rangle_{r,b}$ (or $\langle\sigma\rangle_{E,r,b}$) by an effective cross-section 
\begin{equation}
    \langle\sigma\rangle_{\mathrm{eff},r,b} = (\langle\sigma\rangle_{r,b}, \langle\sigma\rangle_{E,r,b})\prod_i f_{s,i}(N_{\mathrm{col},i}, \dots),
    \label{eq:shielded-xsec}
\end{equation}
where $N_{{\rm col},i}$ is the column densities of species $i$ and the ellipses indicate that shielding functions can also depend on additional user-specified parameters \citep[e.g., the gas Doppler parameter in][]{richings+2014}. As with all other \jaff reaction quantities, shielding functions are full \code{sympy} expressions that can encode arbitrary dependencies. Shielding functions are controlled by user-defined switches, and a set of standard, pre-defined shielding functions are available; we defer a discussion of these to \Cref{sssec:photo-reaction-cross-section}. 

\subsubsection{Radiation-dust coupling}
\label{ssec:dust}

When integrating the radiation field, it is often helpful to include changes due to dust as well as chemical reactions, so that the two can be handled self-consistently within a single integration step. \jaff therefore offers the ability to incorporate dust terms into the equations it generates for the radiation field. When this option is enabled, \jaff modifies \Cref{eq:rad-equation-1,eq:rad-equation-2}, or their energy equivalents \Cref{eq:rad-equation-u-1,eq:rad-equation-u-2}, by adding a term to the radiation evolution equations to account for loss of photons due to dust absorption. These terms take the form   
\begin{align}
    \pdv{n_{\gamma,b}}{t}\Big|_{\rm dust}
    &= - \hat{c}Z_d n_{\mathrm{H,nuc}} n_{\gamma,b}
    \langle\sigma\rangle_{d,b}
    \label{eq:dust-extinction-number}\\
    \pdv{f_{\gamma,b}}{t}\Big|_{\rm dust}
    &= - \hat{c}Z_d n_{\mathrm{H,nuc}} f_{\gamma,b}
    \langle\sigma\rangle_{d,b}
    \label{eq:dust-extinction-flux}
\end{align}
where $Z_d$ is the dust abundance relative to the solar neighborhood, $n_{\mathrm{H,nuc}}$ is the total number density of H nuclei, and $\langle \sigma \rangle_{d,b}$ is the appropriate band-averaged interaction cross-section of radiation with dust per \spe{H} nucleon calculated using \Cref{eq:mean-photo-xsec}. The energy forms are identical except that $n_{\gamma,b}$ and $f_{\gamma,b}$ are replaced by $u_{\gamma,b}$ and $F_{\gamma,b}$, and the number-weighted cross-sections are replaced by energy-weighted ones from \Cref{eq:mean-photo-xsec-energy}.

The reason for using the term interaction is to emphasize the interaction process in focus -- which can be absorption, scattering, extinction, or transport. By default, \jaff assumes the interaction in both \Cref{eq:dust-extinction-number,eq:dust-extinction-flux} to be absorption, but this behavior can be overridden by the user\footnote{Note that \jaff does not model the change in dust temperature due to the different interaction processes; dust temperature calculations are expected to be supplied by the host code.}. \jaff offers a set of built-in choices for the energy-dependent dust cross section $\sigma_d(E_\gamma)$ that enters the averages, as we discuss in \Cref{ssec:multi-format}, but users are free to supply their own as well.

\jaff can also optionally provide a photoelectric heating term in the energy equation to pair with the dust absorption terms, which takes the form
\begin{equation}
    \pdv{e}{t}\Big|_{\rm pe} = \Gamma_{\rm pe}(\chi),
\end{equation}
where $\chi$ is the FUV radiation field normalized to the \citet{draine78} field, and we take the functional dependence $\Gamma_{\rm pe}(\chi)$ from \citet{weingartner-draine-01}; this model is slightly non-linear because it includes the effect of the FUV field on the grain charge state. The default implementation estimates $\chi$ by integrating the radiation field over the $6-13.6$ eV energy range that dominates photoelectric heating; specifically, \jaff calculates $\chi$ as
\begin{equation}
    \chi = \frac{\sum_{b} f_{{\rm PE},b}, u_{\gamma,b}}{u_{\gamma,\rm Draine}},
\end{equation}
where the denominator $u_{\gamma,\rm Draine}$ is the energy density in the \citet{draine78} field integrated over the 6-13.6 eV energy range, and in the numerator the sum is over bands $b$, $u_{\gamma,b}$ is the energy density in band $b$ (computed as $n_{\gamma,b} \overline{E}_{\gamma,b}$ if \jaff is operating in photon number mode), and
\begin{equation}
    f_{{\rm PE},b} = \frac{\int_b \mathcal{W}(E_\gamma) \mathcal{I}_{(6\,{\rm eV},13.6\,{\rm eV})}\, dE_\gamma}{\int_b \mathcal{W}(E_\gamma)\, dE_\gamma}
    \label{eq:fPEb}
\end{equation}
is the fraction of the band that falls into the energy range relevant for photoelectric heating. In \Cref{eq:fPEb}, $\mathcal{I}_{(6\,{\rm eV},13.6\,{\rm eV})}$ is an indicator function that is unity for $E_\gamma \in (6,13.6)$ eV and 0 otherwise.

\subsection{Construction of \jaff networks}
\label{ssec:multi-format}

Because \jaff networks are standard Python objects, and all reaction rates are collections of \code{sympy} expressions, it is possible to construct a network entirely in Python simply by supplying the necessary expressions. However, such manual construction is, as noted in \Cref{sec:intro}, highly error-prone. \jaff therefore provides standard pathways to initialize CRNs, which we describe in this section.

\subsubsection{CRN parsing and processing}
\label{sssec:crn-processing}

\input{tables/supported_reactions}

The non-photochemical parts of the architecture discussed in \Cref{ssec:internal-architecture} can be initialized by reading CRNs written in a number of different formats, including \code{prizmo} \citep{prizmo}, \code{krome} \citep{krome}, \code{udfa} \citep{udfa}, \code{kida} \citep{kida} and \code{uclchem} \citep{uclchem,uclchem-4.0}.\footnote{\code{uclchem} support is still in the alpha phase. \jaff recognizes and computes rates for the standard grain-surface mechanisms (e.g.~freeze-out onto dust grains and thermal, photo- and cosmic-ray-induced desorption) which are parsed using the \code{uclchem} conventions. However, non-standard reaction types, such as those occurring on ices or ones whose rates depend on the rates of other reactions, are not yet supported. Full support for such reactions will be added in a future release.} The full set of supported reaction formats is listed in \autoref{tab:supported_reactions_list}. To accommodate new representation formats that may emerge in the future, \jaff uses a plugin-based system exposing its parsing API to allow for the independent inclusion of parsers without the requirement of updating existing code. The parser architecture allows the parsing of files containing multiple CRN representations via a single input. The \jaff repository includes a set of predefined CRNs which include gas-phase \spe{CO} chemistry \citep{glover2010,glover2011,jappsen_glover_2010,2007ApJ...660.1332J,kingdon_ferland_96,oconnor_2014,kida,udfa}, detailed ISM chemistry \citep{kida_uva_2024}, simplified ISM chemistry \citep{gong17}, primordial chemistry \citep{popsicle}, the \code{umist} CRN \citep{udfa} and a few additional demonstration CRNs. We plan to add more CRNs in the future from various other frameworks.

\jaff parses networks using a two-level strategy to detect and validate reaction lines. The first pass matches an input file against known formats, and the second extracts reactions into  \code{sympy} expressions, with rate coefficients set according to the relevant parser convention or replaced by custom rates (\Cref{ssec:custom-rates}). Once all reactions are parsed, metadata and network properties are either parsed or lazy-loaded for analysis and code generation. 

A key feature of \jaff is that it automates correctness-checking for CRNs, thereby reducing the frequent errors that crop-up in hand-written networks. When it parses a network, \jaff automatically carries out several validity checks, including mass and charge conservation on a per-reaction basis, flagging of duplicate reactions, detection of species that are never produced or never consumed (sinks and sources), identification of isomers sharing an atomic composition, and verification that every cation and anion has a corresponding recombination channel. 

Beyond simple correctness-checking, \jaff provides an extensive library of tools to analyze and visualize chemical networks (see \Cref{fig:species-visualization} for an example). Two networks can be compared reaction-by-reaction and species-by-species. Under the hood, a graph-like, object-oriented data structure exposes an intuitive interface for analyzing and visualizing CRNs and their reaction rates and cross-sections. Production-ready visualization modules and LaTeX-parsing utilities are built into \jaff, alongside drivers for HDF5 and SQLite files that support straightforward table inter-conversion via the \texttt{.toml} configuration specification.

\input{figure_templates/specie_visualization}



\subsubsection{Generation of photochemical rates}
\label{sssec:photo-reaction-cross-section}

The standard CRN formats that \jaff can parse provide flags to indicate which reactions are photon-driven, but as described in \Cref{ssec:photochem}, generating expressions for photochemical reaction rates and corresponding evolution equations for the radiation field requires on-the-fly integration of energy-dependent cross-sections. \jaff automatically carries out these calculations as required using a range of publicly available data.

Photo-reaction cross-sections are broadly divided into three categories -- photoionization, photodissociation and photoabsorption. For photoionization reactions we default to the \href{https://home.strw.leidenuniv.nl/~ewine/photo/}{Leiden database} \citep{hrodmarsson-2023} and fall back to recent high-resolution calculations of cross-sections by \citet{nahar20} and \citet{nh24} which are publicly available in the \href{https://norad.astronomy.osu.edu/}{NORAD database}. Unlike earlier cross-section calculations, the NORAD database calculations include the effects of autoionization resonances, which become important at X-ray energies. In case of photodissociation and photoabsoption, we use the high-resolution photo-dissociation cross-sections from the \href{https://home.strw.leidenuniv.nl/~ewine/photo/}{Leiden database} estimated by \citet{heays17}. The multiplicative shielding functions discussed in \Cref{sssec:shielding} are also estimated from the \href{https://home.strw.leidenuniv.nl/~ewine/photo/}{Leiden database} tables. \jaff supports all reactions and shielding functions present in the \href{https://home.strw.leidenuniv.nl/~ewine/photo/}{Leiden database} for every supported background radiation types.

For dust cross-sections, \jaff uses the data provided by \citet{draine-2003a,draine-2003b,draine-2003c} to estimate the dust extinction, absorption, transport and scattering cross-sections discussed in \Cref{ssec:dust}. All three of the reddening values included in \citeauthor{draine-2003a}'s tabulation,  $R_\mathrm{V}=3.1, 4.0$ and $5.5$, are supported, with the default being 3.1.

\subsubsection{User-defined reactions}
\label{ssec:custom-rates}

Throughout \Cref{ssec:internal-architecture} and \Cref{ssec:photochem}, we have noted that arbitrary reaction rates, energy changes per reaction, heating and cooling expressions, and radiation generation terms can be supplied by the user. While a user can supply these directly in Python, \jaff also defines a network-specific auxiliary file using a custom Python-like syntax format which recognizes \code{sympy} functions, making it possible to store these auxiliary rates alongside the main CRN definition file. An additional advantage of this approach is that it avoids redefinition of custom functions once specified for a CRN. We note that such custom expressions may contain unknown functions, which are kept verbatim in the parsed expression, making it possible to invoke code-specific functions in the generated code. The auxiliary function file can also contain specific directives describing interpolation over data tables, enabling construction of reaction rates or similar expressions that involve table interpolation. These interpolation directives are then translated into code-native interpolation operations at the code generation stage, which we describe in the next section. The following code block provides a demonstration of a custom rate coefficient for \spe{H_2} photodissociation:

\begin{lstlisting}[style=terminal]
@function chemRateCoeff13(chi, av, ncol_H2)
    alpha = 5.7e-11
    gamma = 4.18

    @return alpha * chi * exp(-gamma * av) * fshield_H2(ncol_H2)
\end{lstlisting}

Here, \texttt{chemRateCoeff13} is used to override the default rate for reaction 14 in the network (the indexing convention is zero based); \texttt{chi} and \texttt{av} are symbols which represent the Draine field and visual extinction, respectively, while \texttt{fshield\_H2(ncol\_H2)} is an arbitrary shielding function which takes in the \spe{H_2} column density as an argument. The user can choose to keep this as an unknown function which will be supplied by the simulation code or replace it by one of the shielding prescriptions which come predefined in \jaff \citep[e.g.][]{db96,hg15}.

\subsection{Code generation}
\label{ssec:code-gen}

All the analysis tools discussed above provide \jaff with the unique capability of robust code-generation using \code{sympy}'s code generator -- with support for all major simulation languages like C, C++, FORTRAN, Python, Julia, Rust and R. The generated code can be plugged into practically any codebase. \jaff uses a simple user-friendly templating system, which once defined for a codebase can be reused to generate code to implement any CRN for that codebase. The templating system uses a \jaff-based user friendly syntax to perform substitution, fetching and looping while the output language syntax is automatically detected by \jaff. GPU decorators present in the template are handled seamlessly, making it possible to write templates for GPU-accelerated codes as well as conventional CPU ones. Upon invocation of code-generating functions, the evolution rates, Jacobian or other equations, as per the request of the user/template, are calculated on the fly using \code{sympy} (discussed in \Cref{ssec:internal-architecture,ssec:jacobian,sssec:photo-reaction-rates}).

Currently, \jaff provide predefined templates for \code{flash} \citep{flash}, \code{microphysics} \citep{microphysics} -- the common sub-module used by the \code{castro} \citep{Almgren10a} and \code{maestro} \citep{Fan19a} codes -- and \code{quokka} \citep{quokka, He24a, He24b}, with more to be added in the future. Code generation configurations are controlled by a \texttt{.toml} configuration file which supports features like data table interconversion and table interpolation, providing full support for reactions and heating/cooling mechanisms whose rates are expressed in tabulated form rather than via analytic expressions. Code generation can be done using a simple command-line utility, \jaffgen, specifying which template (i.e., which target code) and which CRN to use:
\\
\begin{lstlisting}[style=terminal]
$ jaffgen --template <template> --network  <network>
\end{lstlisting}

\subsubsection{The right-hand side}

The basic \jaff output is a function in the target language and codebase that takes as input the number densities of each chemical species and the gas internal energy or temperature, and returns as output the derivatives of these quantities appearing on the right-hand side of \Cref{eq:general-rates} and \Cref{eq:general-edot}. In addition to the form given by \Cref{eq:general-edot}, which describes the time evolution of the gas internal energy per unit volume, \jaff can also output the internal energy evolution equation in four alternative forms: specific, molar, per particle, or expressed in terms of a gas temperature. These additional forms are intended to support codes that represent the gas internal energy internally in each of these forms. The equivalent equations available are
\begin{eqnarray}
    \label{eq:dedt-spec}
    \dot{e}_{\rm spec} &=& \frac{\dot{E}}{\rho} \\
    \label{eq:dedt-part}
    \dot{e}_{\rm part} &=& \frac{1}{n_{\rm tot}} \left(\dot{E} - \frac{\dot{n}_{\rm tot}}{n_{\rm tot}} E \right) \\
    \label{eq:dedt-mol}
    \dot{e}_{\rm mol} &=& \frac{N_A}{n_{\rm tot}} \left(\dot{E} - \frac{\dot{n}_{\rm tot}}{n_{\rm tot}} E \right) \\
    \label{eq:dTdt}
    \dot{T} &=& \left(\dot{E} - \sum_i \frac{\partial E}{\partial n_i}\,\dot{n}_i\right)
               \frac{1}{{\partial E}/{\partial T}}
\end{eqnarray}
where subscripts of \emph{$spec$}, \emph{$part$} and \emph{mol} represent specific, per-particle and molar internal energies respectively, $\rho$ is the matter density, $n_\mathrm{tot}$ is the total number density of free particles, and the gas volumetric internal energy is $E$ is computed from the equation of state. By default, \jaff assumes an ideal equation of state  
\begin{equation}
    \label{eq:eos}
    E = \frac{n_\mathrm{tot} k_B T}{(\gamma - 1)}
\end{equation}
where $\gamma$ is the ratio of specific heats. Other equations of state (e.g., for gas which has variable $\gamma$ depending on the chemical composition) are also supported.

For photochemical networks and a particular choice of radiation field energy discretization, this function also provides derivatives of the photon number densities or energy densities, corresponding to the right-hand side of \Cref{eq:rad-equation-1}, including optional terms for dust and the optional auxiliary equation for the radiation flux.

In addition to this primary function, \jaff also emits auxiliary functions for converting from the native state quantities used in the code (most often partial densities of each chemical species, and gas internal energy) to number densities and temperatures. To minimize redundant computation in the generated code, \jaff comes with a switch for common subexpression elimination to all emitted code using \code{sympy}. 

\subsubsection{The Jacobian}
\label{ssec:jacobian}

The set of evolution equations defined in \Cref{ssec:internal-architecture,ssec:photochem} are often insufficient for solving stiff chemical systems due to the restriction on the time step set by the CFL condition \citep{cfl} in explicit solvers. This limitation can be overcome by implicit solvers, which are greatly aided, or sometimes require, a user-supplied Jacobian in addition to the set of coupled evolution equations. Such implicit solvers are adopted by a wide variety of hydrodynamics codes (e.g. \code{quokka} and \code{flash}) and is thus a necessity for any code-generation software. 

Where a Jacobian is requested, \jaff computes it symbolically and on-the-fly by using \code{sympy} to analytically evaluate
\begin{eqnarray}
    \label{eq:jacobian}
    \mathbf{J} = \pdv{\dot{\xi}}{\xi}\ \text{where}\ \xi &\in &  \{n_i, n_{\gamma,b}, f_{\gamma,b}, \mathcal{E}\text{ or }T\} \\
    \text{and }\mathcal{E} &\in & \{ E, e_{\rm spec}, e_{\rm part}, e_{\rm mol} \} \notag
\end{eqnarray}
where $i$ represents the $i^{\text{th}}$ species. In practice, however, rates are generally defined with respect to the gas temperature rather than the specific internal energy. Therefore, for codes using any of the forms of \Cref{eq:dedt-mol,eq:dedt-part,eq:dedt-spec}, \jaff evaluates the $\partial/\partial n_i$ and $\partial/\partial \mathcal{E}$ terms in $\mathbf{J}$ using the chain rule,
\begin{eqnarray}
    \label{eq:jac-ddn-terms}
    \pdv{\dot{\xi}}{n_i}\bigg|_{\mathcal{E}} &=& \pdv{\dot{\xi}}{n_i}\bigg|_T - \pdv{\dot{\xi}}{T}\bigg|_{n_i} 
    \left(\pdv{\mathcal{E}}{n_i}\bigg|_T \bigg/ \pdv{\mathcal{E}}{T}\bigg|_{n_i}\right) \\
    \label{eq:jac-dde-terms}
    \pdv{\dot{\xi}}{\mathcal{E}} &=& \pdv{\dot{\xi}}{T} \pdv{T}{\mathcal{E}} 
\end{eqnarray}
Once \jaff evaluates $\mathbf{J}$ analytically, it then emits a code implementation in the target language and code base.

\subsubsection{ODE solvers}

\jaff supports both codes that include their own-built-in ODE solvers and codes that do not. For codes that do not provide their own ODE solver, in addition to the RHS function and the Jacobian, \jaff can also emit a stiff ODE solver harness that can be compiled alongside the RHS and Jacobian. In this configuration a target code need merely call the \jaff-generated evolution function on each cell or particle in order to advance the chemical (and potentially radiation) state. \jaff currently comes with predefined implementations of a \code{fortran} solver based on \code{dlsodes} \citep{Hindmarsh83a}, a header-only stiff Rosenbrock integrator\footnote{The Rosenbrock integrator provided by \jaff is a slightly modified version of the same integrator supplied by the \code{microphysics} codebase \citep{microphysics} to make it standalone.}  written in C++ (which, being header-only, works equally well for CPU and GPU codes), a Kokkos-based GPU-accelerated ODE solver \citep{Trott22a}, and a harness that uses the Python \code{scipy.integrate.solve\_ivp} solver. \jaff defaults to using \code{dlsodes} for \code{fortran}-based codes, the Rosenbrock solver for C or C++ codes, and \code{solve\_ivp} for Python-based codes.

A key feature of all the solvers that \jaff uses is that they preserve linear invariants \citep{Shampine98a}, meaning that as long as the RHS function for photon absorption and chemical reaction rates is written symmetrically and an exact Jacobian is provided -- both features automatically guaranteed by \jaff's code generation architecture -- conservation to machine precision is guaranteed independent of the accuracy goal used in the ODE solver. Thus for example in a calculation of hydrogen photoionization, the \jaff-generated solver guarantees that the number of ionizations and the number of ionizing photons absorbed match to machine precision. We demonstrate this property in practice in a companion paper (Watt et al., 2026, in preparation) that uses a \jaff-generated network to simulation H~\textsc{ii} region expansion.

While the primary intent of \jaffgen is to produce ready-to-use source code for CRNs that can be inserted into any large hydrodynamics code, the availability of built-in ODE solvers means that \jaff can also produce standalone programs that can be compiled and run by themselves, without the surrounding hydrodynamics architecture. This capability can be used for standalone chemistry calculations.

\section{The test suite}
\label{sec:test-suite}

In the following sections, we demonstrate and verify the accuracy of the code-generation capabilities of \jaff in a variety of scenarios which consist of simple, stiff and radiation-coupled networks. Our goal here is to show that \jaff provides the two major missing capabilities highlighted in \Cref{sec:intro}: automatic conservative code-generation for photochemistry, and the ability to take a single CRN and generate a code implementation of that CRN that can be plugged into sophisticated astrophysical hydrodynamics codes, available in major scientific programming languages, that will allow for scientific reproduction and numerical stability.

\Cref{ssec:h2form,ssec:cie,sssec:gow-oz} demonstrate the latter capability by testing the same CRNs across four codebases - \code{microphysics}, \code{quokka}, \code{flash} and a simple Python initial value problem setup. The first two of these test C++ codes and use the VODE and Rosenbrock ODE solvers respectively, the third tests a \code{fortran} code and uses the \code{dlsodes} solver, and the fourth tests Python code and uses the \code{scipy} \code{solve\_ivp} with the Runge-Kutta 45 explicit integrator. The tests we present in \Cref{sssec:gow-1d-pdr-lpc,sssec:gow-1d-pdr-lpc-hydro,sssec:gow-turbulent-chemistry-box} demonstrate the former capability by using a \jaff-generated network to implement generation of photochemical networks; we also provide additional tests of this capability (e.g.~3D Str\"omgren sphere expansion) along with performance and additional accuracy benchmarks for chemistry, hydrodynamics and radiation against analytical solutions in a companion paper presenting the photochemistry module for \code{quokka} (Watt et al., 2026, in preparation). These tests only use \code{quokka} because among the codes used in our test suite currently only \code{quokka} includes support for radiation transport coupled to photochemistry in its public release\footnote{We do not provide performance benchmarks for other codes because they depend as much on the architecture of the host code as on the \jaff-generated system of ODEs and Jacobian.}.

\subsection{\spe{H_2} formation on dust}
\label{ssec:h2form}

\input{figure_templates/h2form}
\input{figure_templates/h2form_l1}

We start by investigating a toy non-equilibrium CRN where the solution is known analytically: an isothermal gas consisting of pure hydrogen, initially all atomic \spe{H}, which forms \spe{H_2} on the surface of dust grains with a constant rate coefficient $k = 10^{-17}$ cm$^{-3}$ s$^{-1}$ \citep[e.g.,][]{hollenbach-mckee-79}, i.e., this system undergoes the reaction
\begin{equation}
    \label{eq:h2form-reaction}    
    {\rm H} + {\rm H} + grain\rightarrow {\rm H_2} + grain
\end{equation}
The number densities in this system evolve as
\begin{align}
    \label{eq:h2form}
    \pdv{n_\mathrm{H}}{t} = -2k n_\mathrm{H}^2
    \qquad
    \pdv{n_\mathrm{H_2}}{t} = k n_\mathrm{H}^2,
\end{align}
which has the exact solutions
\begin{eqnarray}
    n_{\rm H}(t) &=& \frac{n_{\rm H,0}}{1 + 2 k n_{\rm H,0} t}\\
    n_{\rm H_2}(t) &=& n_{\rm H_2,0} + \frac{n_{\rm H,0}-n_{\rm H}}{2}
\end{eqnarray}
where $n_{\rm H,0}$ and $n_{\rm H_2,0}$ are the number density of \spe{H} and \spe{H_2} at $t=0$, respectively. We implement a \jaff CRN containing this reaction, use it to generate implementations of the network in \code{flash}, \code{microphysics}, \code{quokka}, and \code{python\_ivp}, and then evolve the chemical state in a single cell, disabling hydrodynamics and heating and cooling. We show the results in \Cref{fig:h2form}, demonstrating that all code bases produce essentially-identical results that agree with the exact solution. We quantify the level agreement with the analytical solution using the $L_1$ error, defined as
\begin{equation}
    \label{eq:l1-error}    
    L_1 = |y_{\rm n}-y_{\rm a}|
\end{equation}
where subscripts ${\rm n}$ and ${\rm a}$ represent the numerical and analytical quantities, respectively. For this test, we set the integrator absolute tolerance to $10^{-20}$ and vary the relative tolerance $r_{\rm tol}$ from $10^{-1}$ to $10^{-12}$.   eWe demonstrate the results for the different integrators in \Cref{fig:h2-l1-errors}. The $L_1$ errors come out proportional to $r_{\rm tol}$ in most cases. The best results are produced by the fifth order Rosenbrock integrator which produces $L_1$ errors of the order of $10^{-12}$ even for high values of $r_{\rm tol}$. 

\subsection{Collisional ionization of \spe{H}}
\label{ssec:cie}

\input{figure_templates/cie}

Next, we demonstrate the thermodynamic capabilities of \jaff using a CRN which consists only of endothermic reactions. Specifically, we implement a CRN containing the following hydrogen collisional and recombination reactions
\begin{eqnarray}
    \label{eq:cie-reactions}
    \spe{H} + \spe{e^-} & \rightarrow & \spe{H^+} + \spe{e^-} + \spe{e^-} \\
    \spe{H^+} + \spe{e^-} & \rightarrow & \spe{H}
\end{eqnarray}
We take the collisional ionization and recombination cooling and reaction rates for these reactions from equations 12a, 13a, 26a  and 27a of \citet{cen1992} respectively. For the sake of limiting the cooling rate in the absence of any heating mechanism, we omit collisional line cooling and bremsstrahlung cooling. We initialize the simulation with a gas temperature $T = 15000$ K and H$^+$, $e^{-}$, and H abundances of $10^{-3}$, $10^{-3}$, and $1-10^{-3}$ per H nucleon, respectively. We again use a single zone for this test, with an initial \spe{e^{-}} and \spe{H^{+}} number density of $10^{-3}\ \rm cm^{-3}$, with no hydrodynamics, and simulate the system for $10^{16}$ s.

We show species abundances, gas internal energy, and gas temperature as a function of time in \Cref{fig:cie}, for two cases: one where we include the change in gas internal energy due to endothermic reactions, and one where we ignore the chemical cooling terms. Note that in the latter case the gas internal energy is constant, but the gas is not isothermal, since the temperature still decreases due to the increase in free particles over time as the neutral hydrogen ionizes, and \jaff takes this effect into account.

We see that all the codes again produce near-identical results, demonstrating that \jaff~evolves the thermodynamics consistently in each codebase; differences are at the $< 1\%$ level, which is expected given that we use an accuracy goal of 1\% in the ODE solver. If we disable the change in internal energy due to endothermic reactions (open points in \Cref{fig:cie}), the gas remains at constant internal energy, and equilibrates to a state where the ionization and recombination rates are equal. By contrast, if we include chemical cooling (solid points), the gas initially begins to ionize, but then it cools as ionizations followed by recombinations remove energy from the system, and the ionization fraction again declines. In both cases the behavior is as expected.

\subsection{\spe{H}-\spe{He}-\spe{C}-\spe{O} interstellar chemistry}
\label{ssec:gow}

Our third test uses a real-life stiff chemistry network that has been used extensively in simulations: the H-He-C-O network CRN proposed by \citet[GOW17 hereafter]{gong17}. This network serves as a strong test case for \jaff's robustness since it contains 50 reactions, 18 species and comes with custom rate coefficients, complex internal energy evolution mechanisms, dependence on the external radiation field, and the requirement for table interpolation. This test also allows us compare the \jaff-generated results against the independently-generated results of \citetalias{gong17} which uses the SUNDIALS \citep{sundials} CVODE solver in BDF mode.

\subsubsection{Single-zone tests with fixed radiation fields}
\label{sssec:gow-oz}

\input{tables/gow-oz-initial-conditions}
\input{figure_templates/gow-oz-chi0.1}
\input{figure_templates/gow-oz-chi1}
\input{figure_templates/gow-oz-chi10}

We start with one-zone tests with fixed radiation fields using \jaff-generated versions of the \citetalias{gong17} network implemented in \code{quokka}, \code{flash}, \code{microphysics}, and \code{solve\_ivp}. Among these, \code{quokka} and \code{microphysics} evolve the gas specific internal energy (\Cref{eq:dedt-spec}) while \code{flash} and \code{solve\_ivp} evolve the temperature (\Cref{eq:dTdt}). In these tests, we initialize single computational cells with a logarithmically spaced range of number densities ranging from $10$ to $1000$ \spe{H} nuclei per $\mathrm{cm}^3$, initial chemical abundances adopted from \citetalias{gong17} (\Cref{tab:gow-initial-conditions}), and initial temperatures of 100 K; we also adopt \citetalias{gong17}'s default Solar neighborhood dust-to-gas ratio. In the single-zone tests we do not perform any radiative transfer calculations, and instead impose a fixed, spatially uniform cosmic ray ionization rate and radiation field. We use a primary cosmic ray ionization rate per hydrogen nuclei $\zeta = 2\times 10^{-16}$ s$^{-1}$, and far ultraviolet radiation fields $\chi=0.1$, $1$ and $10$ measured in \citet{draine78} units. We run the simulations using the \jaff-generated network to evolve the species abundances and gas temperature, and with hydrodynamics disabled, until the system reaches equilibrium. In all cases, we use a relative tolerance of $10^{-12}$ and an absolute tolerance of $10^{-20}$. We define our accuracy goal as $\pm0.1\%$ of the publicly available implementation of \citetalias{gong17} for the quantities being compared. We do not require exact agreement, since slight differences in the implementation of the chemical network and the numerical integration methods may lead to small discrepancies between the different implementations.

We show the results of this experiment in \Cref{fig:gow-oz-chi01,fig:gow-oz-chi1,fig:gow-oz-chi10} for $\chi = 0.1, 1,$ and 10, respectively. The left columns show species abundances (top row) and temperatures (bottom row) as a function of density, comparing the four code results (points) to the \citetalias{gong17} reference solution (solid lines), and the right two columns show the PDF (computed from a gaussian kernel density estimate) and CDF of the relative differences between the \jaff-computed solutions and the reference solution for species abundances and temperatures across all cells and chemical species; for the purpose of these plots we define the relative error as $\epsilon_{\rm X}=|1 - y_{\rm X}^{\rm code}/y_{X}^{\rm Gong+ 2017}|$, where $y_{\rm X}^{\rm code}$ is the result of quantity $\rm X$ generated by each code running the \jaff-generated network, and $y_{\rm X}^{\rm Gong+2017}$ is the corresponding result generated by \citetalias{gong17}'s publicly available implementation. The principal conclusion to take from the plots is that the \jaff-generated results in all cases agree with the reference solution within the integrator tolerance, in most cases substantially better, with a median error (denoted by dashed vertical lines) of $10^{-14}-10^{-15}$. This demonstrates one of \jaff's key capabilities: that it can encode a chemical network in a way that enables generation of runnable code that will produce the same answer to within solver tolerances for any codebase, independent of the type of rate equation and language, while running on CPU or GPU.

\subsubsection{1D PDR with live photochemistry}
\label{sssec:gow-1d-pdr-lpc}

\input{figure_templates/pdr_bands_n100_chi1}
\input{figure_templates/pdr_physics_n100_chi1}

Our next goal is to demonstrate how a \jaff generated network performs in live-photochemistry simulations. Once again, we evolve the CRN prescribed by \citetalias{gong17}, but this time we use version in \code{quokka} that includes live photochemistry; for this purpose we allow \jaff to generate all the photochemical reaction rates automatically, via the pathway described in \Cref{sssec:photo-reaction-rates}, rather than relying on the hard-coded functions expressed in terms of a \citet{draine78} radiation field as in the original \citetalias{gong17} network. For the dust extinction calculations discussed in \Cref{ssec:dust}, we use the $R_{\rm V} = 3.1$ curve discussed in \Cref{sssec:photo-reaction-cross-section}. For these calculations we use a simulation domain of $8192\times1\times1$ linearly spaced grid cells, and we disable hydrodynamics. The boundary conditions are periodic in the $y$ and $z$ directions, outflow on the $+x$ face of the domain, and inflow on the $-x$ face, enabling us to inject a radiation flux into the domain. We add two additional adaptive mesh refinement levels immediately adjacent to the $-x$ boundary to provide better resolution of the low-optical depth region close to the boundary. All simulations begin with the same uniform gas density and chemical abundances used in \Cref{sssec:gow-oz}.

We consider three separate cases for the tests. For the first we use one radiation band from $6-13.6\text{ eV}$, treating the entire far ultraviolet as a single band. In the second we use two bands -- $6 - 11.2$ and $11.2-13.6 \text{ eV}$ -- to separate the FUV into a lower band that drives photoelectric heating but cannot dissociate H$_2$ and an upper band that covers the Lyman-Werner bands and thus captures H$_2$-dissociating photons. Furthermore, we note that the \citetalias{gong17} model and models without self consistent radiative transfer, in general, use dust extinction coefficients estimated accurately using only the interaction range of the spectrum for the species under consideration. In this regard, our dust extinction model discussed in \Cref{ssec:dust} is an approximation when the band lengths are large. Therefore, we include a third run with 4 radiation bands -- $6-8.6$, $8.6-11.2$, $11.2-12.4$ and $12.4-13.6\text{ eV}$. This allows us to evaluate how reducing the band widths affects the accuracy of our results. For each of these three cases, we carry out a run with a $\chi=1$ injected radiation field strength at the $-x$ boundary -- meaning that we set the radiation energy density and flux at the boundary by integrating a \citet{draine78} over the bands we use -- and an initial of $n_{\rm H}=100\ {\rm cm}^{-3}$. We also add one additional, control run without live radiation, where we use the same incident radiation and shielding prescription as \citetalias{gong17}; the purpose of this run is simply to verify that, when we drop our self-consistent treatment of radiative transfer and photochemistry, we recover results identical to those of \citetalias{gong17}, so any differences we find in the simulations with self-consistent radiative transfer and photochemistry are due to that.

We show the results in \Cref{fig:scm_n100_chi1}, plotting the abundance of every species and the gas temperature and radiation energy density as a function of column density from the $-x$ face of the domain. We can first note that our control test passes: the open points, which show the \jaff-generated results without self-consistent radiation transfer and photochemistry, sit exactly on top of the gray line, which shows the \citetalias{gong17} reference solution, verifying that the underlying chemistry network is the same. On the other hand, the blue, orange and pink lines, which show the one-band, two-band and four-band solutions with self-consistent radiative transfer and photochemistry, show small but clear differences. The column density to the PDR front (where the H/H$_2$ transition occurs) in the two-band and four-band runs differ by a quarter to half a dex against the reference, while they have a mutual difference of $\sim0.2$ dex i.e. the results have nearly converged with the decrease in the band length.

The difference arises primarily from the treatment of dust extinction in the photo-electric heating band: \citetalias{gong17} uses a hard-coded order of magnitude estimate of $10^{-21}\,\mathrm{cm}^2$ motivated by \code{despotic} \citep{despotic} and \citet{weingartner-draine-01}. While this is a reasonable approximation in a treatment that averages the entire FUV into a single band, it neglects the considerable (factor of $\approx 2$) difference in dust cross section over the $6-13.6$ eV energy range, something that is calculated self-consistently by \jaff as discussed in \Cref{ssec:dust}. The importance of this effect is evident from the final panel of \Cref{fig:scm_n100_chi1}, which shows the considerable difference in radiation energy density between the two bands. The one-band \jaff case is closer to the \citetalias{gong17} solution, but this at least in part due to an accidental cancellation between two approximations -- the dust extinction cross-section order of magnitude approximation which overestimates the position of the front towards a relatively lower column density and the use of gray radiation, which averages out most of the physics in the $11.2-13.6\,\mathrm{eV}$ band, in turn underestimates the position of the front towards relatively higher column density.
\subsubsection{1D PDR with live photochemistry and hydrodynamics}
\label{sssec:gow-1d-pdr-lpc-hydro}

In our next test, we demonstrate all the pieces enabled by \jaff in a single test: live hydrodynamics, radiative transfer, and photochemistry all solved simultaneously using \code{quokka}. While simulations which include the coupled time evolution of all three ingredients have been available for some time in the case of simplified photoionization chemistry networks, this test represents the first application of a similar methodology to the relatively more complex chemistry occurring in a PDR. The simulation setup and initial conditions for this test are identical to the two band cases discussed in the previous section, but in this case we do not disable hydrodynamics, and allow the system to evolve hydrodynamically over time. For the hydrodynamic part of the simulation we impose first order extrapolating boundary conditions on the $-x$ face of the simulation domain. 

We show the results of this run with the solid green line in \Cref{fig:scm_n100_chi1-hydro}. Since this runs is not expected to reach equilibrium in our simple 1D setup, we evolve the setup till $3\times10^{15}\ {\rm s}$, which is the same time required for the non-hydrodynamic runs in \Cref{sssec:gow-1d-pdr-lpc} to reach equilibrium. In this case, the PDR front is a steep non-stationary front which advances in the $+x$ direction with time. This effect arises due the high pressure created in the upstream region where the photo-electric and photo-dissociation heating leads to rapid expansion of the gas. Use of the \jaff-generated photochemistry network allows us to capture all of these effects.

\subsubsection{3D turbulent chemistry box}
\label{sssec:gow-turbulent-chemistry-box}

So far, we have only focused on one zone and 1D tests; in contrast, current numerical simulations using hydrodynamics are mostly three dimensional. Therefore, for our final test, we choose a 3D turbulence box setup using a \jaff-generated chemistry network to mimic the environment of a molecular cloud.  In this setup, the computational domain is initialized with a resolution of $512^3$ spanning $50\ \rm pc$ on each side with periodic boundary conditions. Once again, we use the ISM chemical network prescription by \citetalias{gong17} with $\chi=1$. However, unlike our previous setups which explicitly calculated the column density of each species, we use the capped Jeans length based local shielding prescription by \citet{Safranek-Shrader17a}\footnote{See \citet{gaches+2025b} for further justification.}; we do not use radiative transfer in this test. Our initial hydrodynamic conditions are motivated by \citet{micic+2012} -- gas with a hydrogen nuclei number density of $30\ \rm cm^{-3}$ at a temperature of $100\ \rm K$ embedded in a medium driven by purely solenoidal turbulence forcing maintaining a target velocity dispersion of $2$ km s$^{-1}$; our turbulence driving implementation uses the driving module of \citet{Federrath22a}. We run our setup for two turbulent crossing times ($t_{\rm turb}$) with pure hydrodynamics to allow the velocity dispersion to reach a steady state, following which we enable chemistry and evolve the setup for another four $t_{\rm turb}$. \Cref{fig:turbbox} demonstrates our results at $t=2t_{\rm turb}$ after chemistry has been enabled. The top and middle panels represent 2D field maps of different quantities at the $z=25\ \rm pc$ face; the bottom left and middle plots show the probability density function of the primary \spe{H} and \spe{C} species against the hydrogen nuclei number density while the bottom right panel shows the temperature phase plot. Once again, we see that the setup produces physically expected results -- molecular species such as \spe{H_2} and \spe{CO} are dominant in high density clumps with their median in the range of $10^2-10^3\ \rm cm^{-3}$. On the other hand, ionized species like \spe{H^+} and \spe{C^+} survive primarily in diffuse regions; their median occurs at $\sim 10$ and $40\ \rm cm^{-3}$, respectively. The gas temperature ranges from $\sim10-2000\ \rm K$; the gas is hotter in diffuse regions, while dense regions are cool molecular clumps. Additionally, we note a kink in the temperature phase plot in regions where \spe{C^+} transition to \spe{C} (bottom-right and middle panels); this effect arises due to reduction in \spe{C^+} line cooling which is eventually compensated by cooling from \spe{C} and \spe{CO}. Thus, a \jaff generated network correctly preserves the physics for higher dimensional setups using stiff chemical networks.
\input{figure_templates/turbulent_chemsitry_box}

\section{Summary}
\label{sec:summary}

In this work we present \jaff\ -- an extensible, cross-platform astrochemistry framework written in Python and built on the symbolic-algebra capabilities of \code{sympy}. It parses all major astrochemical network formats and provides real-time network diagnostics, validation, analysis and visualization. From a single command, its templated, configuration-driven code generator emits simulation code in seven programming languages that can be plugged into practically any codebase. In addition, \jaff offers native support for self-consistent multi-band photochemistry and dust extinction.

We have validated the generated code with a test suite spanning simple toy networks and a real-world interstellar photochemistry network across a range of physical and simulation scenarios -- single-zone models, 1D photo-dissociation regions with pure chemistry run across multiple codebases in different languages and running on both CPU and GPU, and multi-band radiation-coupled chemistry with and without hydrodynamics. Across these tests, \jaff correctly generates the rate equations and Jacobian terms including custom rates, heating and cooling terms consisting of data-table interpolation functions and reproduces the expected results.

The capability offered by \jaff opens up exciting prospects for future applications. First, it can upgrade a hydrodynamics code to a coupled chemistry-hydrodynamics code, similarly it can upgrade a radiation-hydrodynamics code to a photochemistry-radiation-hydrodynamics code. 

Second, \jaff enables clean experiments comparing codes that differ in their treatments of hydrodynamics and radiative transfer by ensuring that treatments of chemistry can be made identical and thus that the codes are solving the same underlying equations -- a problem that has proven consistently challenging even in code comparison projects where the chemical networks are exceedingly simple \citep[e.g.,][]{Iliev09a} and has proven near-impossible for more complex photochemistry networks \citep[e.g.,][]{Rollig07a}.

Third, by lowering the considerable barrier involved in hand-coding a complex chemical network in a low-level language such as C or \code{fortran}, \jaff will stimulate further development of new approximate chemistry networks for different applications. \jaff greatly reduces the time required to implement and test new networks in a live simulation, speeding the development cycle and allowing faster experimentation.

\section*{Data availability}

\jaff is available under an open source license at \href{https://github.com/jaff-chemistry/jaff}{github.com/jaff-chemistry/jaff}. Detailed documentation is provided at \href{https://jaff-chemistry.github.io/jaff/}{jaff-chemistry.github.io/jaff}. The simulation data used in this article are publicly available on Zenodo at \url{https://doi.org/10.5281/zenodo.24638212}.

\section*{Acknowledgments}

The authors wish to thank Leiden University's Lorentz Center and the participants of the Lorentz Center workshop ``Numerical Recipes in Star Formation" (July 2025) for providing a highly collaborative and interactive environment that fostered the development of this work. AS acknowledges discussions with Christoph Federrath on setting up a \jaff template for \code{flash}. AS, JW, CCH, and MRK acknowledge funding from the Australian Research Council through Laureate Fellowship FL220100020. This research was supported by the Australian Government's National Collaborative Research Infrastructure Strategy (NCRIS), with access to computational resources provided by the Pawsey Supercomputing Centre and the National Computational Infrastructure (NCI Australia) through the National Computational Merit Allocation Scheme, award jh2. 

\bibliographystyle{aasjournal}

\bibliography{bibliography}




\end{document}

%% file: header/imports.tex
\usepackage{xcolor}
\usepackage{textgreek}
\usepackage[utf8]{inputenc}
\usepackage[english]{babel}
\usepackage{hyperref}
\usepackage{color,colortbl}
\usepackage{tensind}
\usepackage{verbatim}
\usepackage[normalem]{ulem}
\usepackage{orcidlink}
\usepackage{soul}
\usepackage{amsmath}
\usepackage{booktabs}
\usepackage{xspace}
\usepackage{physics}
\usepackage{cleveref}
\usepackage{listings}

%% file: header/setup.tex
\hypersetup{
    unicode,
    colorlinks=true,
    linkcolor=linkcolor,
    citecolor=linkcolor,
    filecolor=linkcolor,
    urlcolor=linkcolor,
}
\definecolor{linkcolor}{rgb}{0.0,0.3,0.5}
\tensordelimiter{?}
\DeclareGraphicsExtensions{.bmp,.png,.jpg,.pdf}
\lstdefinestyle{terminal}{
    backgroundcolor=\color{gray!10},   
    basicstyle=\small\ttfamily,        
    breaklines=true,                   
    frame=single,                      
    frameround=tttt,                   
    framesep=5pt,                      
    rulecolor=\color{gray!30},         
    keywordstyle=\color{blue},         
    language=bash                      
}

\graphicspath{ {./figures/} }
\defcitealias{gong17}{GOW17}

\addto\extrasenglish{%
}

%% file: header/macros.tex
\newcommand{\jaff}{\textsc{jaff}\xspace}
\newcommand{\code}[1]{\textsc{#1}}
\newcommand{\jaffgen}{\texttt{jaffgen}\xspace}
\newcommand{\jaffx}{\texttt{jaffx}\xspace}
\newcommand{\spe}[1]{\ensuremath{\mathrm{#1}}}

%% file: figure_templates/arch.tex
\begin{figure}
    \centering
    \includegraphics[width=\columnwidth]{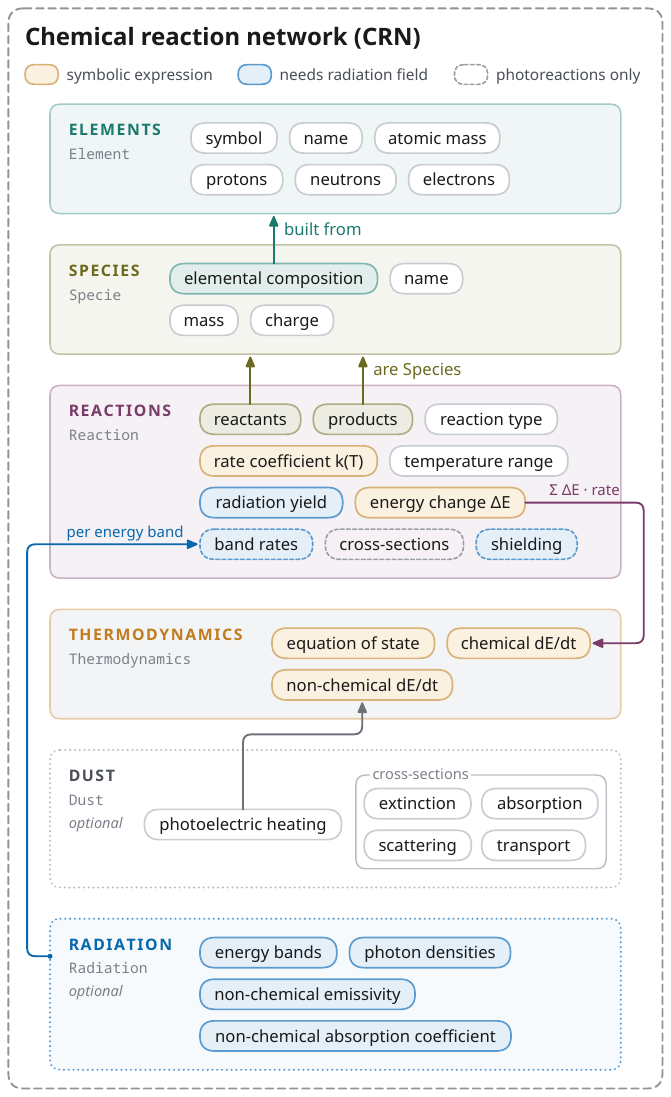}
    \caption{Simplified schematic of \jaff's CRN representation architecture. The green, olive and pink components represent the elements, species and reactions catalog respectively. Pills in yellow are symbolic expressions while blue components indicate that they are active when using a discretized radiation field prescription.\\}
    \label{fig:network-arch}
\end{figure}

%% file: figure_templates/oh_cross_section.tex
\begin{figure}
    \centering
    \includegraphics[width=\columnwidth]{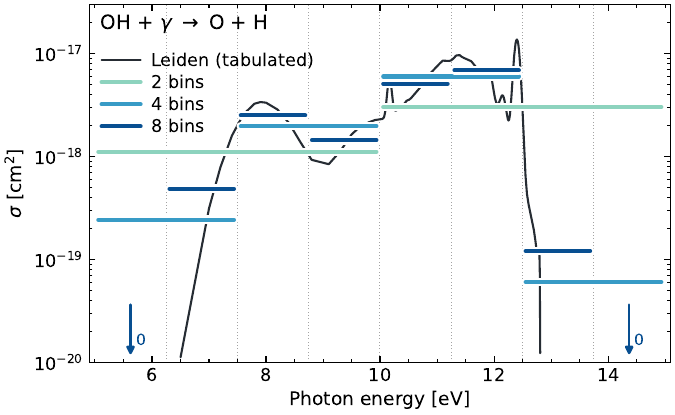}
    \caption{Example calculation of band-averaged cross-sections for \spe{OH} photodissociation from \jaff. The black line shows the frequency-dependent cross-section for the reaction $\mathrm{OH} + \gamma \to \mathrm{O} + \mathrm{H}$, taken from the Leiden database \citep{hrodmarsson-2023}. Cyan, light blue and and dark blue lines denote the band averaged cross-section $\langle\sigma\rangle_b$ computed by \jaff~for a discretization of the energy range $5.5 - 14.5$ eV into two, four and eight bands, respectively while arrows indicate $\langle\sigma\rangle_b=0$. The eight band case is the closet representation of the actual spectrum while the the two band case is least accurate.\\}
    \label{fig:oh_xsec}
\end{figure}

%% file: tables/supported_reactions.tex
\begin{table*}
\footnotesize
\renewcommand{\arraystretch}{1.4}
\begin{tabular*}{\textwidth}{@{\extracolsep{\fill}}lll}
\toprule
Reaction & Form & Rate coefficient \\
\midrule
\textbf{Shared} & & \\
Two-body gas-phase & $\mathrm{A + B \rightarrow C + D}$ & $\alpha\,(T/300)^{\beta}\exp(-\gamma/T)$ \\
Three-body & $\mathrm{A + B + M \rightarrow AB + M}$ & Same as two-body reactions \\
Cosmic-ray ionisation & $\mathrm{A} + \mbox{cosmic ray} \rightarrow \mathrm{A^{+} + e^{-}}$ & $\alpha\zeta$, \; $\alpha\zeta/\zeta_0$ \\
Cosmic-ray-induced photoreaction & $\mathrm{AB} + h\nu_{\rm CR} \rightarrow \mathrm{A + B}$ & $\alpha\gamma\,(T/300)^{\beta}\,\zeta/(1-\omega)$, \;  \\
Ultraviolet photoreaction & $\mathrm{AB} + h\nu \rightarrow \mathrm{A + B}$ & $\alpha\chi\,e^{-\gamma A_V}$ \\
 & & $\int 4\pi\,\sigma(E)\,J(E)\,E^{-1}\,dE$ \\
\midrule
\textbf{KIDA} & & \\
Ion--neutral & $\mathrm{A^{+} + B \rightarrow C^{+} + D}$ & $\alpha\beta\,[0.62 + 0.4767\,\gamma\,(300/T)^{1/2}]$ \\
 & & $\alpha\beta\,[1 + 0.0967\,\gamma\,(300/T)^{1/2} + 300\,\gamma^{2}/(10.526\,T)]$ \\
Cosmic-ray-proton ionisation & $\mathrm{A} + \mbox{cosmic ray} \rightarrow \mathrm{A^{+} + e^{-}}$ & $2\alpha\zeta\,n(\mathrm{H_2})/n_{\rm H}$ \\
\midrule
\textbf{UCLCHEM} & & \\
Freeze-out & $\mathrm{A \rightarrow A_{surface}}$ & $[1 + \beta\,1.671\times10^{-3}/(T a)]\,v_{\rm th}\,\sigma_{\rm gr}\,(T/m)^{1/2}$ \\
Cosmic-ray desorption & $\mathrm{A_{surface} \rightarrow A}$ & $4\pi \times 1.64\times10^{-4} \times 4\,\alpha\,\sigma_{\rm gr}\,\phi\,\zeta$ \\
Cosmic-ray photodesorption & $\mathrm{A_{surface}} + h\nu_{\rm CR} \rightarrow \mathrm{A}$ & $4.875\times10^{3}\,\alpha\,\sigma_{\rm gr}\,Y\zeta\,[1 + \chi\,e^{-1.8A_V}/(\epsilon\zeta)]$ \\
\bottomrule
\end{tabular*}
\caption{\protect\raggedright Standard reaction forms recognised by the \jaff network parsers and their rate coefficients. Forms used by more than one format are listed first; forms unique to one format follow under that format's heading.
\protect\\ \textit{Rate parameters:} $\alpha$, $\beta$, $\gamma$: parameters of the network line; $T$: gas temperature (K).
\protect\\ \textit{Radiation:} $\zeta$: cosmic-ray ionisation rate (s$^{-1}$); $\zeta_0 = 1.36\times10^{-17}$~s$^{-1}$: reference rate; $\chi$: ultraviolet field strength; $A_V$: visual extinction; $h\nu_{\rm CR}$: cosmic-ray-induced ultraviolet photon; $\sigma(E)$: photo cross-section; $J(E)$: mean intensity.
\protect\\ \textit{Species:} M: third body; $n(\mathrm{H_2})$, $n_{\rm H}$: H$_2$ and total hydrogen-nucleus number densities; $m$: species mass.
\protect\\ \textit{Grains:} $a$: grain radius; $\omega = 0.5$: grain albedo; $v_{\rm th}$: thermal velocity prefactor; $\sigma_{\rm gr}$: cross-section per H nucleus; $\phi$: molecules desorbed per cosmic-ray impact; $Y$: ultraviolet photodesorption yield; $\epsilon$: cosmic-ray ultraviolet efficiency.
}
\label{tab:supported_reactions_list}
\end{table*}

%% file: figure_templates/specie_visualization.tex
\begin{figure}
    \centering
    \includegraphics[width=\columnwidth]{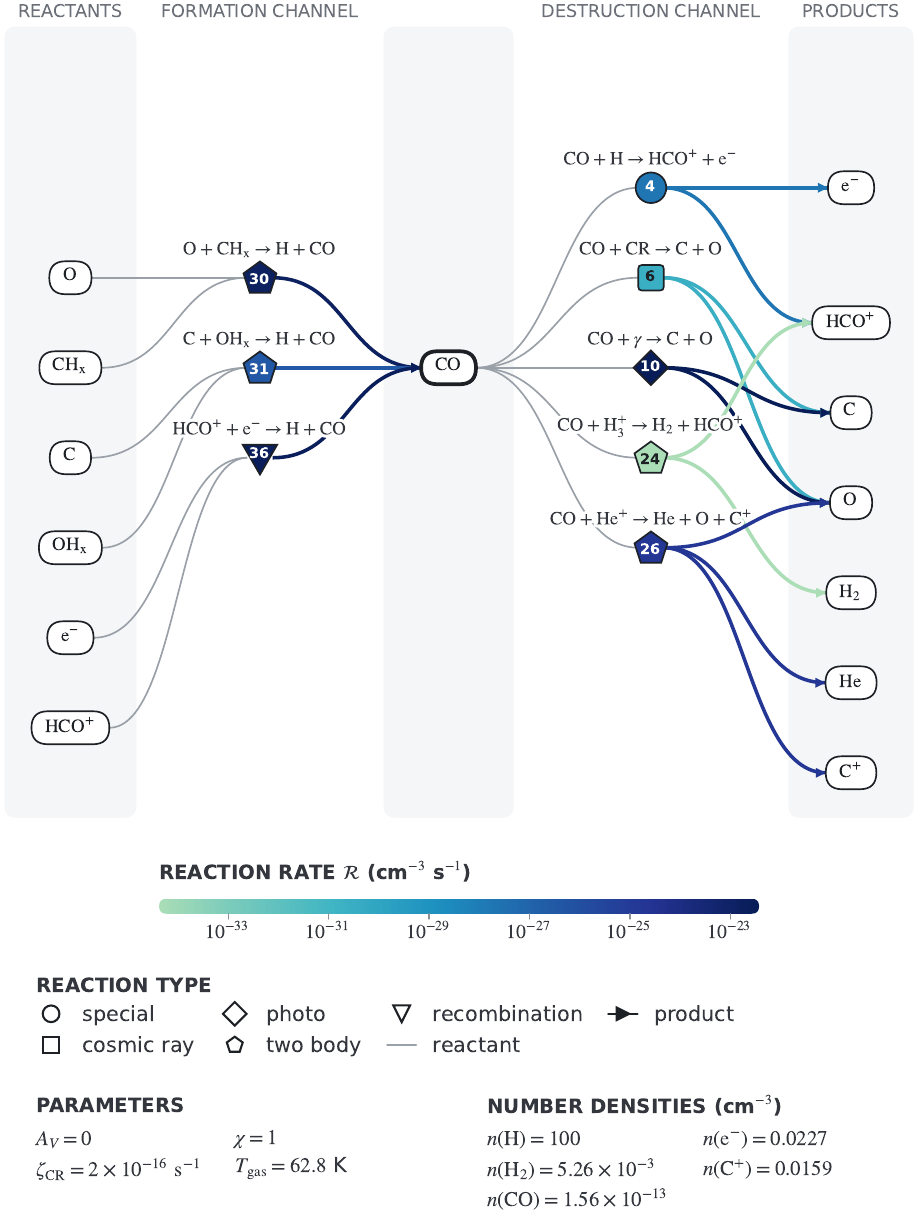}
    \caption{Visual representation of the formation and destruction channels of \spe{CO} in the \citet{gong17} CRN generated by the \texttt{Network.visualize.species("CO")} method. Colored lines indicate the direction of the reaction while the color denotes which rate dominates.\\}
    \label{fig:species-visualization}
\end{figure}

%% file: figure_templates/h2form.tex
\begin{figure}
    \centering
    \includegraphics[width=\columnwidth]{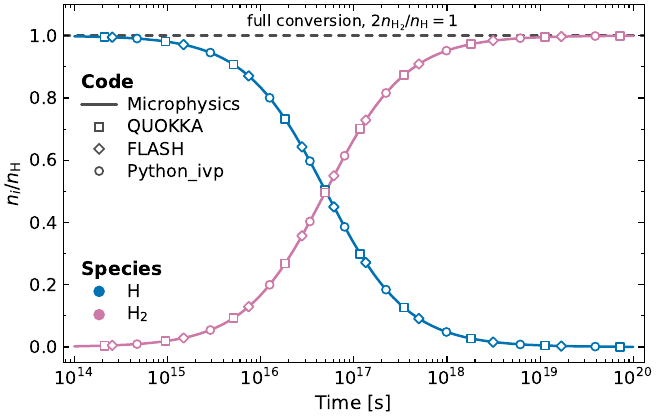}
    \caption{Demonstration of \jaff generated code for an isothermal hydrogen formation network (discussed in \Cref{ssec:h2form}) in four codebases -- \code{flash}, \code{microphysics}, \code{quokka} and \code{python\_ivp}. The \spe{H_2} and \spe{H} abundances for all codes converges to the analytical solution (represented by the violet dashed line) which is equal to the initial abundance of \spe{H} when all \spe{H} atoms have been converted to \spe{H_2} molecules.\\}
    \label{fig:h2form}
\end{figure}

%% file: figure_templates/h2form_l1.tex
\begin{figure}
    \centering
    \includegraphics[width=\columnwidth]{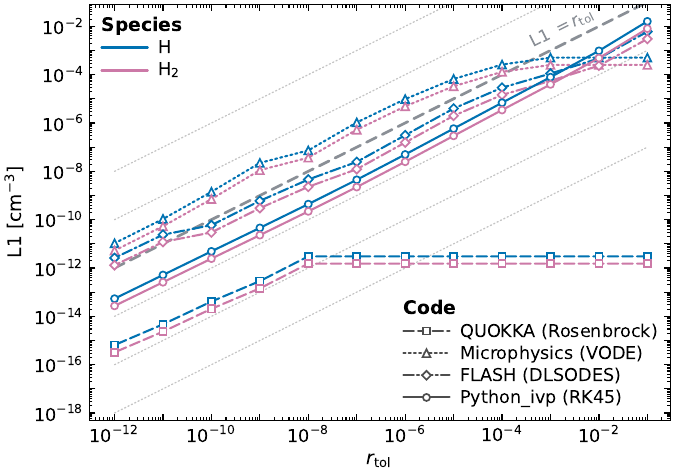}
    \caption{$L_1$ errors as a function of $r_{\rm tol}$ for the various integrators used for the \spe{H_2} formation test. The dashed line denote the slope of the $L_{1}=r_{\rm tol}$ curve while the dotted lines act as parallel guide lines.\\}
    \label{fig:h2-l1-errors}
\end{figure}

%% file: figure_templates/cie.tex
\begin{figure}
    \centering
    \includegraphics[width=\columnwidth]{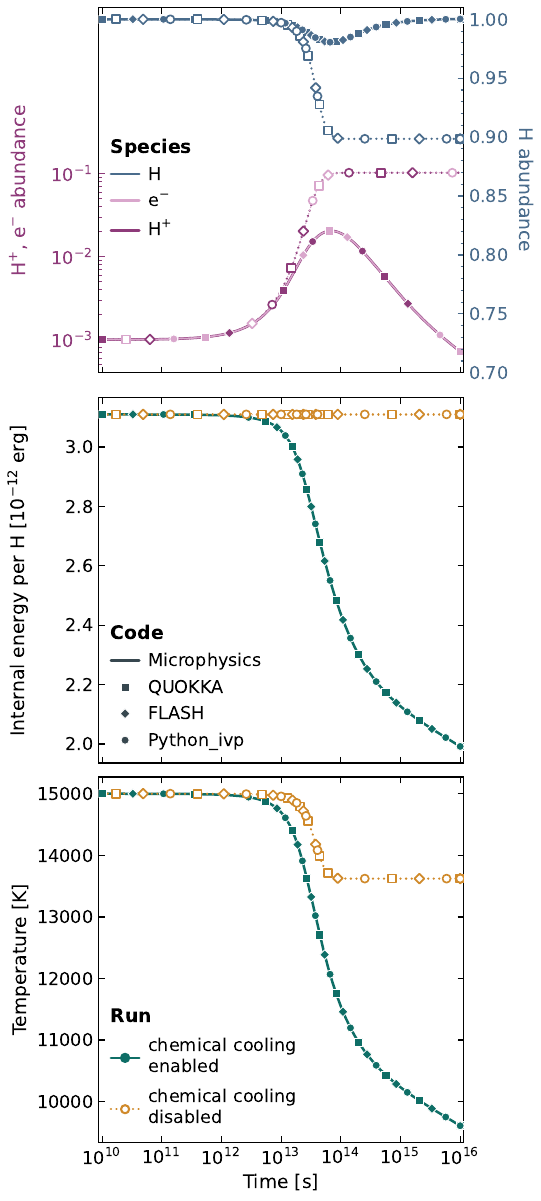}
    \caption{Demonstration of \jaff generated code for a collisional ionization network (\Cref{ssec:cie}). Panels show, from top to bottom, species abundances (\spe{e^-} and \spe{H^+} abundance follow the logarithmic axis on the left, while \spe{H} abundance follows the linear axis on the right), internal energy per \spe{H} nucleon, and temperature. Different symbols show results produced using \jaff-generated networks in different codebases -- \code{flash}, \code{microphysics}, \code{quokka} and \code{python\_ivp} -- as indicated in the legend. Solid points and lines represent the runs with chemical cooling enabled in \jaff while hollow points with dashed lines represent runs where chemical cooling is disabled.}
    \label{fig:cie}
\end{figure}

%% file: tables/gow-oz-initial-conditions.tex
\begingroup
    \setlength{\tabcolsep}{4pt}
    \renewcommand{\arraystretch}{1.2}
    \setlength\extrarowheight{2pt}
    \begin{table}
        \centering
        \begin{tabular}{ c c c c }
            \toprule
            Species & Abundance & Species & Abundance \\
            \midrule
            \spe{H}   & $1.0$                   & \spe{CO}  & $1.0\times10^{-7}$ \\
            \spe{H^+}  & $1\times10^{-4}$        & \spe{HCO^+} & $1\times10^{-40}$ \\
            \spe{e^-}  & $2.60283\times10^{-4}$  & \spe{O}   & $3.2\times10^{-4}$ \\
            \spe{H_2}  & $1\times10^{-10}$       & \spe{Si}  & $1.7\times10^{-6}$ \\
            \spe{H_2^+} & $1\times10^{-40}$       & \spe{Si^+} & $1\times10^{-40}$ \\
            \spe{He}  & $1\times10^{-1}$        & ${\rm CH}_x$  & $1\times10^{-40}$ \\
            \spe{He^+} & $1.450654\times10^{-8}$ & ${\rm OH}_x$  & $1\times10^{-40}$ \\
            \spe{C}   & $1\times10^{-40}$       & \spe{H_3^+} & $2.681411\times10^{-7}$ \\
            \spe{C^+}  & $1.6\times10^{-4}$      & \spe{O^+}  & $1\times10^{-40}$ \\
            \bottomrule
        \end{tabular}
        \caption{\protect\raggedright Initial abundances for the H-He-C-O network tests, expressed as number of each species per H nucleon. The values are motivated by \citetalias{gong17}.\\}
        \label{tab:gow-initial-conditions}
    \end{table}
\endgroup

%% file: figure_templates/gow-oz-chi0.1.tex
\begin{figure*}
    \centering
    \includegraphics[width=\textwidth]{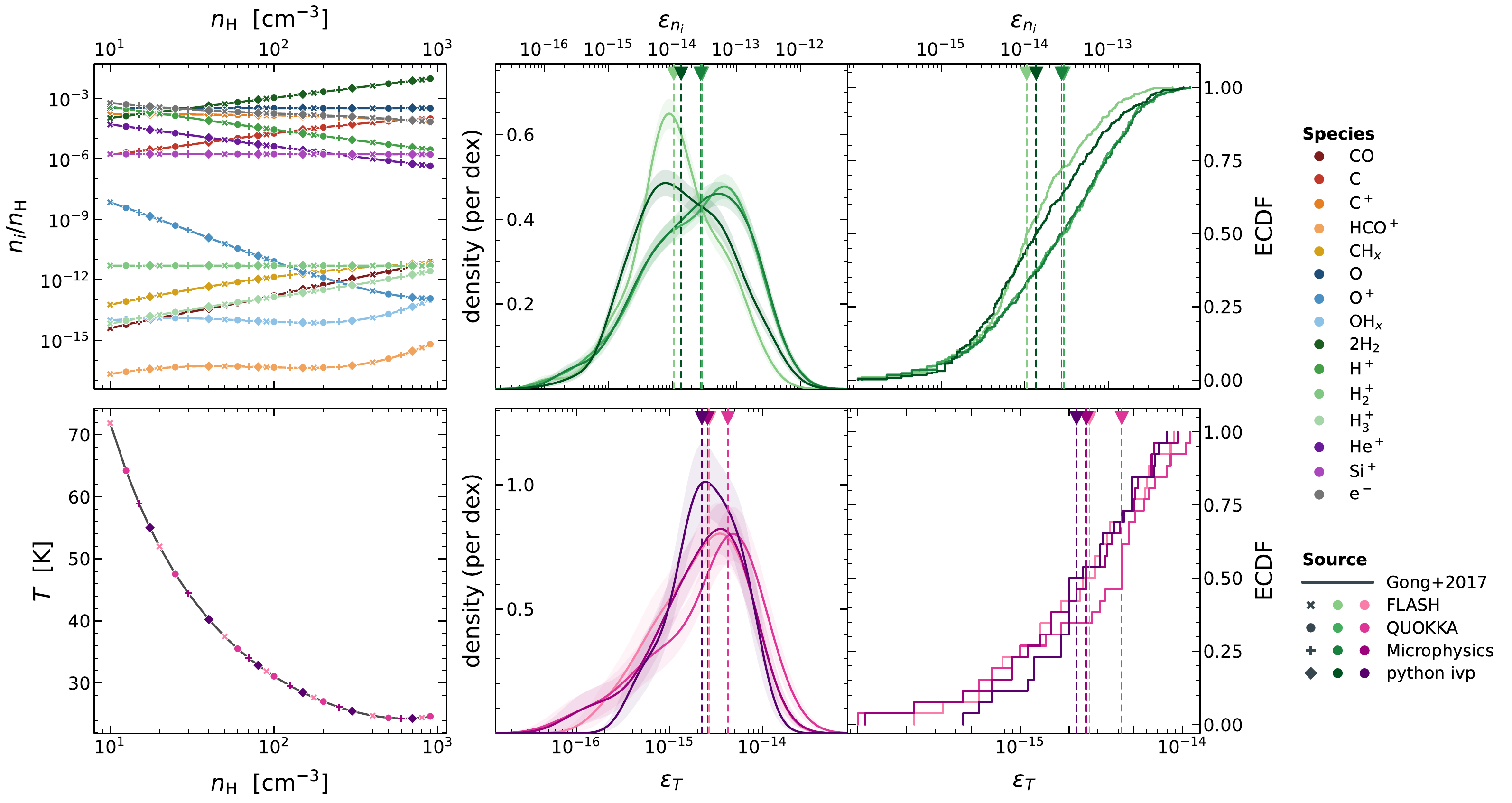}
    \caption{Demonstration of \jaff generated code for the \citetalias{gong17} ISM network in four code bases -- \code{flash}, \code{microphysics}, \code{quokka} and \code{python\_ivp} at an incident radiation field intensity of $\chi=0.1$. In the left column, the solid lines show the reference results for species abundances (top) and gas temperature (bottom) computed by \citetalias{gong17}'s publicly-available implementation, while points show the results generated by each of our four test codes. The middle and right columns show the PDF and inverse CDF of the relative differences $\epsilon$ between the four codes and the reference solution; see main text for the exact definition of the relative difference that we use. Colored bands represent the standard deviation with respect to the PDF while the median error is indicated by the dashed lines.\\}
    \label{fig:gow-oz-chi01}
\end{figure*}

%% file: figure_templates/gow-oz-chi1.tex
\begin{figure*}
    \centering
    \includegraphics[width=\textwidth]{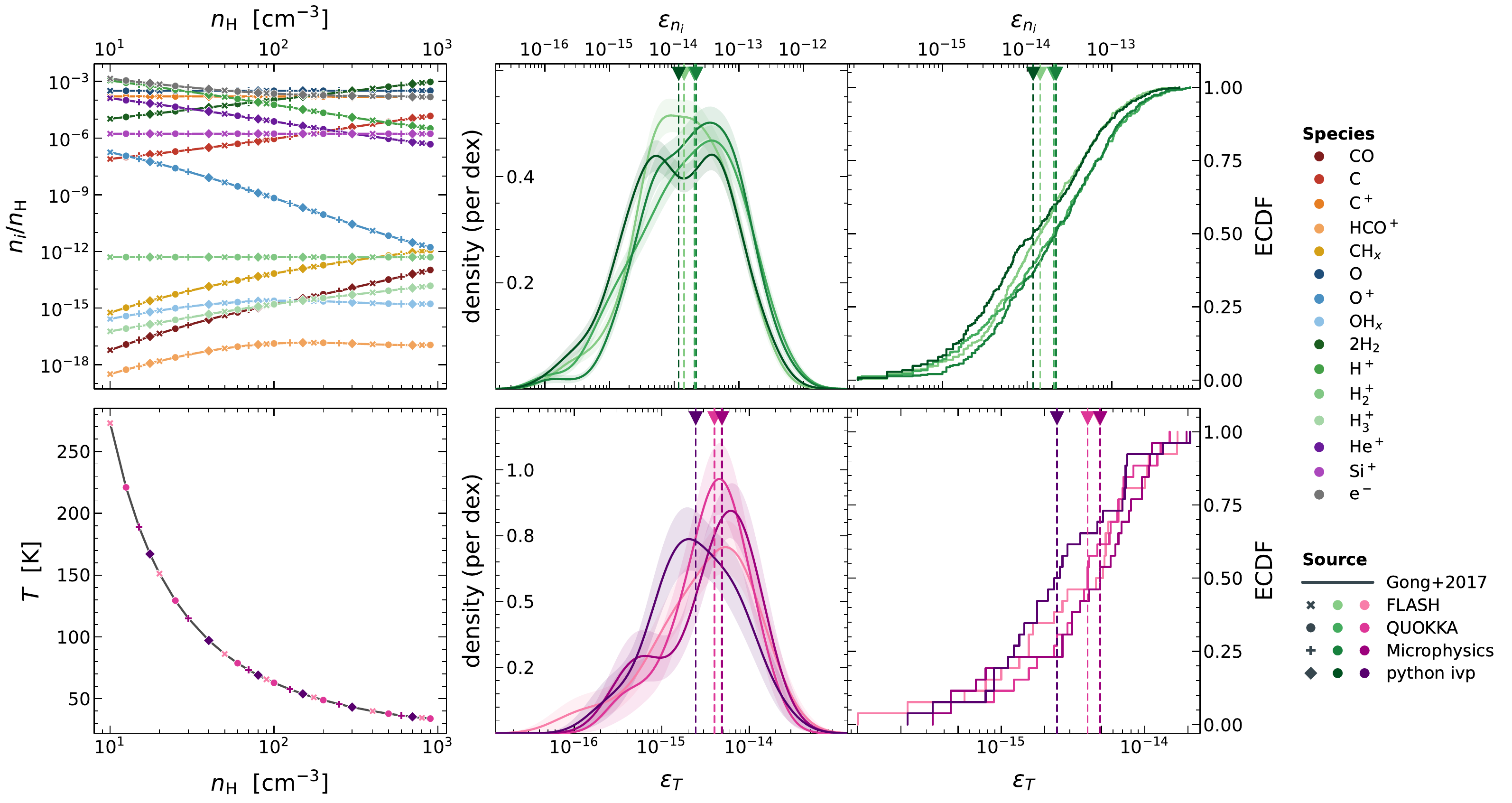}
    \caption{Same as \Cref{fig:gow-oz-chi01}, but for $\chi = 1$.\\}
    \label{fig:gow-oz-chi1}
\end{figure*}

%% file: figure_templates/gow-oz-chi10.tex
\begin{figure*}
    \centering
    \includegraphics[width=\textwidth]{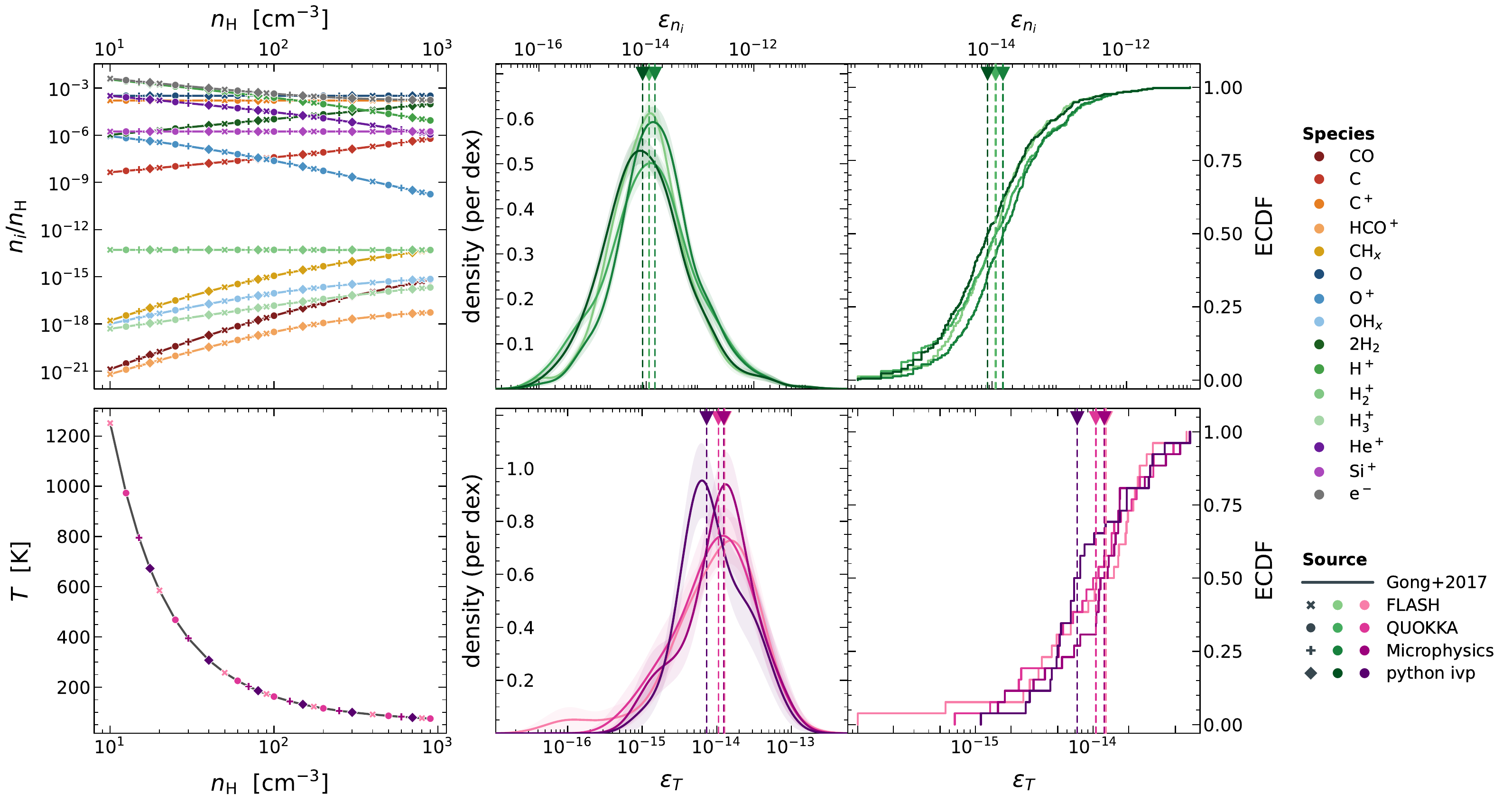}
    \caption{Same as \Cref{fig:gow-oz-chi01}, but for $\chi = 10$.\\}
    \label{fig:gow-oz-chi10}
\end{figure*}

%% file: figure_templates/pdr_bands_n100_chi1.tex
\begin{figure*}
    \centering
    \includegraphics[width=0.97\textwidth]{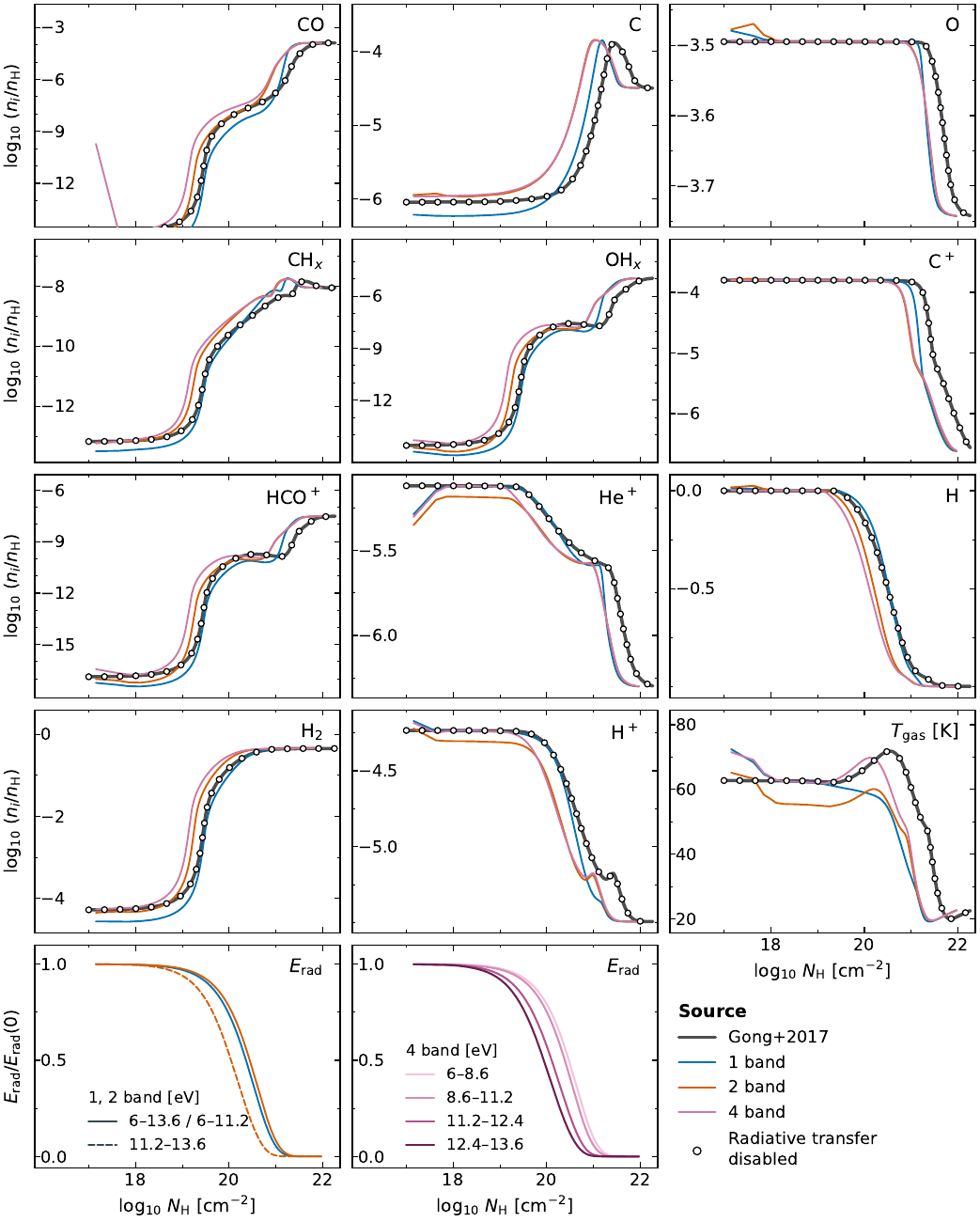}
    \caption{Species abundances, gas temperature, and radiation energy density normalized to the value at the left edge of the simulation domain as a function of gas column density from the domain left edge for our tests with live photochemistry. Gray lines show the \citetalias{gong17} solution computed without live photochemistry, and open circles show the \jaff-generated results for the control run where we have turned off live radiation and hydrodynamics. Blue, orange and pink lines show the results generated by \jaff using self-consistent, live photochemistry and radiative transfer using one, two and 4 radiation bands, respectively. Sharp changes for $N_{\rm H}\leq10^{18}\ \rm cm^{-2}$ are boundary artifacts arising due to low resolution; \citetalias{gong17} uses a logarithmically spaced grid while we use linear spacing; as a result, the resolution is poor at the $-x$ boundary and adding more AMR levels produces spurious oscillations.}
    \label{fig:scm_n100_chi1}
\end{figure*}

%% file: figure_templates/pdr_physics_n100_chi1.tex
\begin{figure*}
    \centering
    \includegraphics[width=\textwidth]{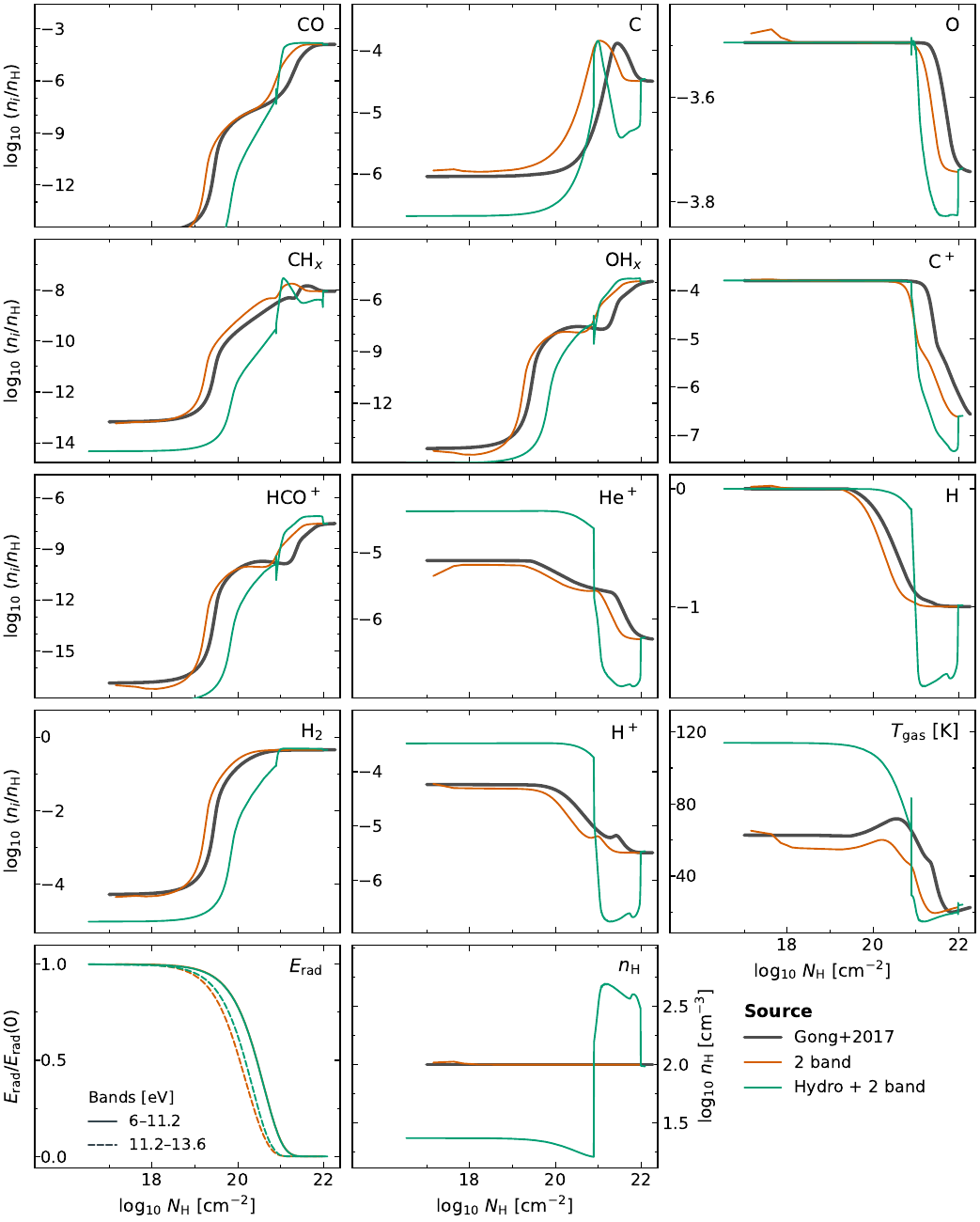}
    \caption{Species abundances, gas temperature, hydrogen nuclei number density, and radiation energy density normalized to the value at the left edge of the simulation domain as a function of gas column density from the domain left edge at $t=3\times10^{15}\ \rm s$ for our tests with self-consistent photochemistry. Gray lines show the \citetalias{gong17} solution computed without self-consistent photochemistry. Orange lines show the results generated by \jaff using self-consistent photochemistry and radiative transfer using two radiation bands. Green lines show a result that combines self-consistent photochemistry and radiative transfer with hydrodynamics. The numerical oscillations are the same boundary artifacts of \Cref{fig:scm_n100_chi1} which have been advected inward.}
    \label{fig:scm_n100_chi1-hydro}
\end{figure*}

%% file: figure_templates/turbulent_chemsitry_box.tex
\begin{figure*}
    \centering
    \includegraphics[width=\textwidth]{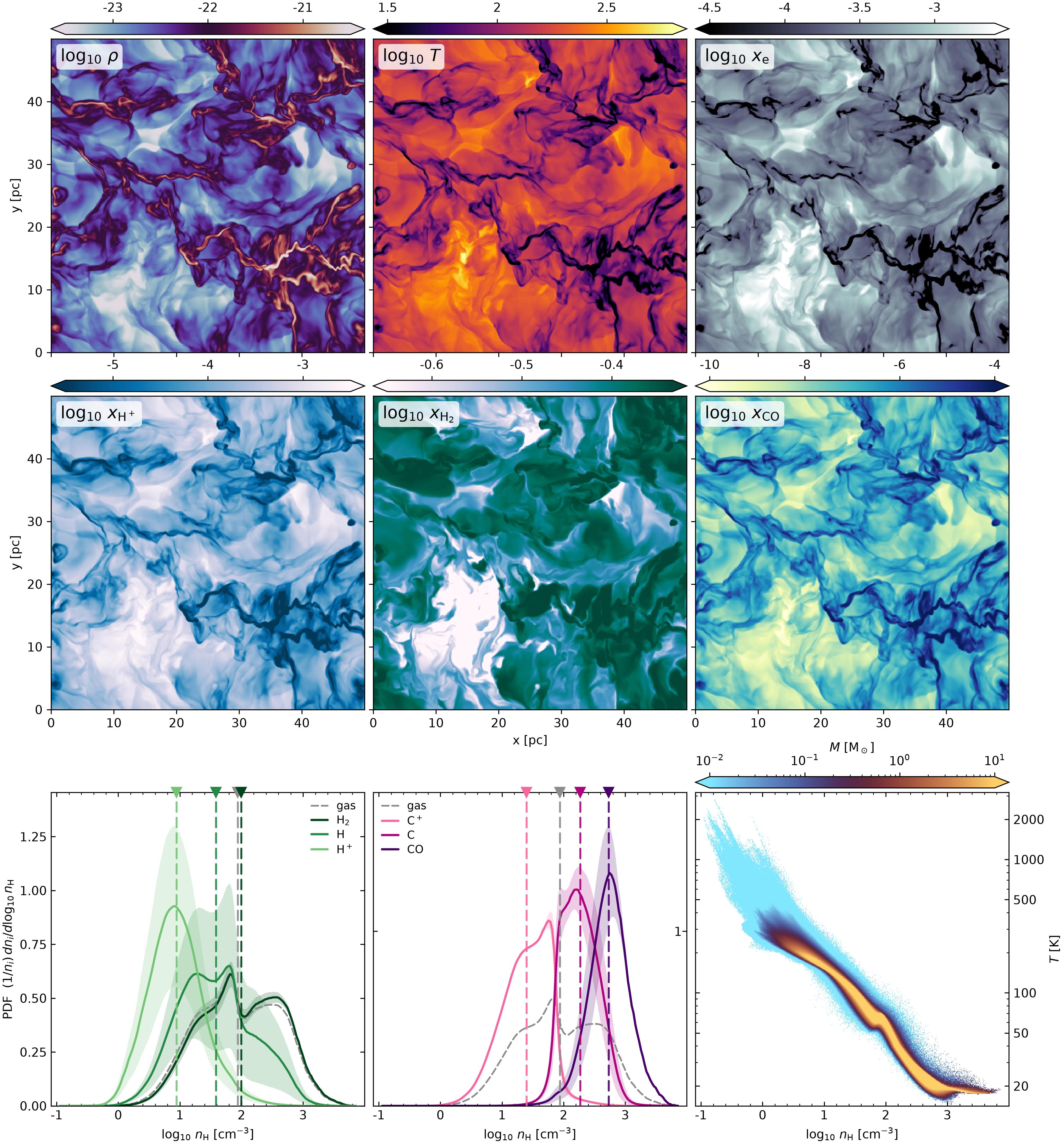}
    \caption{Top and middle panel shows contour plots of gas density, temperature and number densities of \spe{e^-}, \spe{H^+}, \spe{H_2} and \spe{CO} from left to right respectively; densities are in units of g cm$^{-3}$, temperatures are in K, and abundances $x$ are dimensionless. The bottom left plot shows the PDFs of \spe{H_2}, \spe{H} and \spe{H^+} density, along with the total mass density PDF (dashed curve), and the bottom middle plot shows the same for \spe{CO}, \spe{C} and \spe{C^+}. Colored bands indicate the standard deviation in that logarithmic bin and dashed vertical lines indicate the median. The bottom right plot shows the 2D temperature-density phase diagram.}
    \label{fig:turbbox}
\end{figure*}